%% file: 1_frontiers-main.tex
\documentclass[utf8]{FrontiersinHarvard} 

\usepackage{url,hyperref,lineno,microtype,subcaption}
\usepackage[onehalfspacing]{setspace}
\usepackage{3_preamble}

\def\keyFont{\fontsize{8}{11}\helveticabold }
\def\firstAuthorLast{Bong {et~al.}} 
\def\Authors{Heejong Bong\,$^{1}$, Valérie Ventura\,$^{2,6,7}$, Eric A. Yttri\,$^{4,6,7}$, Matthew A. Smith\,$^{5,6,7}$ and Robert E. Kass\,$^{2,3,6,7,*}$}
\def\Address{$^{1}$Department of Statistics, Purdue University, West Lafayette, IN, USA \\
$^{2}$Department of Statistics and Data Sciences, Carnegie Mellon University, Pittsburgh, PA, USA \\
$^{3}$Machine Learning Department, Carnegie Mellon University, Pittsburgh, PA, USA \\
$^{4}$Department of Biological Sciences, Carnegie Mellon University, Pittsburgh, PA, USA \\
$^{5}$Department of Biomedical Engineering, Carnegie Mellon University, Pittsburgh, PA, USA \\
$^{6}$Neuroscience Institute, Carnegie Mellon University, Pittsburgh, PA, USA \\
$^{7}$Center for the Neural Basis of Cognition, Carnegie Mellon University, Pittsburgh, PA, USA
}
\def\corrAuthor{Robert E. Kass}

\def\corrEmail{kass@stat.cmu.edu}

\begin{document}
\onecolumn
\firstpage{1}

\title[Latent Dynamic Analysis via Sparse Banded Graph]{Cross-Population Amplitude Coupling in High-Dimensional Oscillatory Neural Time Series} 

\author[\firstAuthorLast ]{\Authors} 
\address{} 
\correspondance{} 

\extraAuth{}

\maketitle

\begin{abstract}

Neural oscillations have long been considered important markers of interaction across brain regions, yet identifying coordinated oscillatory activity from high-dimensional multiple-electrode recordings remains challenging. We sought to quantify time-varying covariation of oscillatory amplitudes across two brain regions, during a memory task, based on local field potentials recorded from 96 electrodes in each region. We extended Canonical Correlation Analysis (CCA) to multiple time series through the cross-correlation of latent time series. This, however, introduces a large number of possible lead-lag cross-correlations across the two regions. To manage that high dimensionality we developed rigorous statistical procedures aimed at finding a small number of dominant lead-lag effects. The method correctly identified ground truth structure in realistic simulation-based settings. When we used it to analyze local field potentials recorded from prefrontal cortex and visual area V4 we obtained highly plausible results. The new statistical methodology could also be applied to other slowly-varying high-dimensional time series.

\tiny
 \keyFont{ \section{Keywords:} latent factor models, high-dimensional time-series, multiset CCA, cross-region dynamic connectivity} 
\end{abstract}

\input{sections/1_introduction}

\input{sections/2_methods}

\input{sections/3_results}
\input{sections/4_discussions}

\section*{Conflict of Interest Statement}

The authors declare that the research was conducted in the absence of any commercial or financial relationships that could be construed as a potential conflict of interest.

\section*{Author Contributions}

Bong, Ventura, Yttri and Kass conceived and designed the research; Bong and Ventura performed the simulation studies; Yttri and Smith performed experiments; Bong implemented the algorithm and analyzed the data; Bong, Yttri, Smith and Kass interpreted results of the experiments; Bong prepared the figures; Bong, Ventura and Kass drafted the manuscript; Bong, Ventura, Yttri, Smith and Kass edited and revised the manuscript; Bong, Ventura, Yttri, Smith and Kass approved the final version of the manuscript.

\section*{Funding}
Bong, Ventura and Kass are supported in part by NIMH grant R01 MH064537. Yttri is supported by NIH grant (1R21EY029441-01) and the Whitehall Foundation. Smith is supported by NIH (R01EY022928, R01MH118929, R01EB026953, P30EY008098) and NSF (NCS 1734901) grants.



\section*{Data Availability Statement}
The datasets analyzed for this study can be found in \citet{snyder2022utah}: \url{https://kilthub.cmu.edu/articles/dataset/Utah_array_recordings_from_visual_cortical_area_V4_and_prefrontal_cortex_with_simultaneous_EEG/19248827}.

\bibliographystyle{Frontiers-Harvard} 
\bibliography{2_refs}

\newpage
\appendix
\renewcommand{\thesection}{S\arabic{section}}
\renewcommand{\thefigure}{S\arabic{figure}}
\renewcommand{\thetable}{S\arabic{table}}
\renewcommand{\thealgorithm}{S\arabic{algorithm}}

\setcounter{figure}{0}
\setcounter{table}{0}
\setcounter{algorithm}{0}

\section*{Supplementary Material}

\input{appendices/a_proofs}

\clearpage
\input{appendices/b_algorithm}

\clearpage
\input{appendices/c_supp_simulation}

\clearpage
\input{appendices/d_supp_experiment}

\end{document}

%% file: sections/1_introduction.tex
\section{Introduction}\label{sec:introduction}

\begin{figure}[t!]
  \includegraphics[scale=0.6]{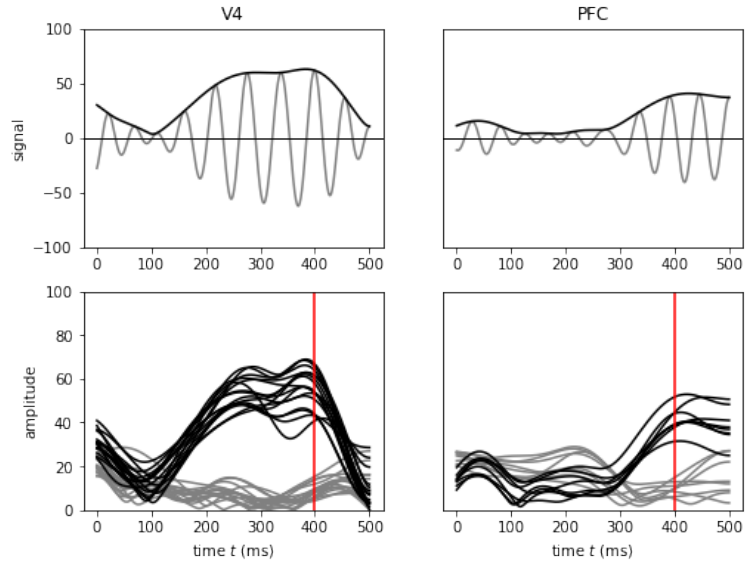}
    
  \caption{\sl {\bf Bandpass-filtered LFP amplitude envelopes.} The top panels display beta-range filtered LFPs (gray) together with their amplitude envelopes (black)  from a pair of electrodes in V4 (left) and PFC (right) on a single trial, during the delay period of a working memory task. (The filtering is described in \cref{sec:simulation_estimation}.) The bottom panels display amplitude envelopes of active electrodes in the two brain regions for two trials, those for one trial in black and those for the other trial in gray. The beta amplitude envelopes show a consistent temporal pattern within each region, for both trials, and, at $400$ ms (red vertical line), the two groups of amplitidue envelopes from the two brain regions illustrate correlated behavior across the two trials in the sense that the black amplitude curves for the first trial are elevated, compared with the gray for the second trial, in both V4 and PFC. This correlation, however, varies across time.}
  \label{fig:intro}
\end{figure}

Contemporary technologies for recording neural activity can produce multiple time series in each of two or more brain regions simultaneously \citep[e.g.,][]{jun2017fully,steinmetz2018challenges}, enabling identification of cross-regional interactions relevant to cognitive processes that support behavior. {One such process is working memory \citep{miller2018wm2,schmidt2019beta}.} During a variety of experimental working memory tasks, strong neural oscillations in the beta range (16-30 Hz) have been observed, leading to the proposal that beta oscillations serve to coordinate activity across regions \citep{miller2018wm2}. To identify coordinated oscillatory activity, many statistical methods based on coherence, or phase coupling, have been developed, studied, and applied \citep{belluscio2012cross, klein2020torus, ombao2022spectral, urban2023oscillating}. As shown in \Cref{fig:intro}, however, oscillatory amplitudes (amplitude envelopes) also vary substantially and may well exhibit statistical dependence across regions. Our aim is to detect whether these {slowly-varying} amplitude envelopes tend to fluctuate together, and to pinpoint when during a task such coupling occurs. This form of amplitude-based coordination reflects a different aspect of neural communication than phase-based measures and requires statistical tools that can accommodate smooth, non-oscillatory signal components. In this paper, we develop methods for analyzing groups of {slowly-varying multiple time series, and apply them} to uncover coordinated amplitude activity between brain regions.



A common strategy for quantifying inter-areal association is to reduce multichannel recordings within each brain region to a small number of region-level summaries and then compute pairwise association measures between regions. For oscillatory neural signals, this is often done by computing amplitude envelope correlations after extracting instantaneous amplitudes via the Hilbert transform, based on region-level summaries \citep{bruns2000amplitude,zamm2018amplitude,raghavan2024gamma}. Constructing these summaries typically requires additional preprocessing, such as selecting a single representative channel per region \citep{bruns2000amplitude}, applying spatial filters \citep{zamm2018amplitude}, or designating channels based on expert judgment \citep{raghavan2024gamma}.

A similar dimension reduction principle underlies methods developed for non-oscillatory recordings such as fMRI. Time Delay (TD) analysis \citep{mitra2015lag,mitra2018spontaneous,raut2019organization} averages voxel-level signals within each region to obtain a single regional time-series. It then compares pairs of regional time=series by measuring how much one signal is delayed relative to the other using cross-covariance, resulting in a single lag value for each region pair that can be further analyzed using PCA. Cyclicity analysis \citep{shahsavarani2020comparing,abraham2024hemodynamic} instead aggregates directionality measures across voxel-level interactions to produce a singed scalar measure that represents region-level summary that captures region-level lead-lag relationships.


These approaches collapse multichannel activity into signals that mix task-relevant and task-irrelevant processes, 
which reduces sensitivity to between-region coupling and obscures the subpopulations that carry the interaction. We instead developed a method to learn low-dimensional and time-localized components directly from multichannel data, without spatial priors. By maximizing between-region dependence (relative to within-region variance) our approach increases sensitivity to inter-areal communication.

The data that motivated this work are local field potentials (LFPs; \citealp{buzsaki2012origin,pesaran2018investigating}), recorded from prefrontal cortex (PFC) and visual area V4 during repeated trials of a working memory task \citep{khanna2019dynamic}.
PFC is generally considered to exert control aimed at areas involved in perceptual processing \citep{miller2001integrative}, such as V4, which is active during the retention of higher-order visual information (e.g. color and shape) and attention to visual objects \citep{orban2008higher,fries2001modulation}. Both regions exhibit strong oscillatory activity during visual working memory tasks, and their interaction, particularly through oscillatory coupling, has been implicated in supporting working memory function \citep{sarnthein1998synchronization,liebe2012theta,miller2018wm2,gregoriou2009high}.

The data consist of 96 LFP time series for each brain region (192 time series in total), offering millisecond temporal precision, and are relatively precise spatially, as well. These signals reflect neural activity across large populations of neurons within each region and exhibit substantial spatial correlation. To summarize the cross-regional interactions more succinctly, dimensionality reduction becomes a natural and effective step \citep{gallego2022local}. If we were to analyze the data at a single time point, we would have repeated pairs of 96-dimensional observations, for which canonical correlation analysis (CCA) would be a standard tool to assess statistical dependence. Here we build on a latent variable formulation of CCA known as probabilistic CCA (pCCA), and extend it to handle temporally structured data, as illustrated in \Cref{fig:pCCA}(b). 
In the extended model described in \cref{sec:pCCAtimeseries}, each of the two high-dimensional time series is driven by a latent univariate time series, with the correlation between the two latent time series representing the statistical dependence we seek to identify (see \Cref{fig:pCCA_over_time}). \cref{thm:equivalence_to_genvar} demonstrates that this model yields estimates equivalent to those from a generalization of classical CCA, called multiset CCA \citep{kettenring1971canonical}). This model-based framework allows for flexible modeling of covariance structure and facilitates principled high-dimensional inference.

Related latent variable methods include Gaussian Process Factor Analysis (GPFA, \citealp{lakshmanan2015extracting,semedo2020statistical,yu2009gaussian}) and state-space models \citep{goris2014partitioning}. GPFA assumes that latent time series follow a low-dimensional Gaussian process, while state-space models typically impose autoregressive dynamics. Although these methods were originally devised to capture shared low-dimensional structure within a single brain region, they can be embedded in a two-step approach to discover cross-regional interaction: (1) apply those methods separately to each brain region, and (2) compute cross-correlations between the resulting latent trajectories. While common in neuroscience, this strategy may lack statistical power when within-region activity does not align with cross-regional communication. For example, in recordings from macaque visual areas V1 and V2, \citet{semedo2019cortical} found that communication occurred in a dedicated ``communication subspace'' distinct from the dominant within-region activity. Ignoring this distinction can lead to underestimating cross-regional coupling and reduced sensitivity \citep{semedo2020statistical}.

To address this, alternative methods have been proposed. \citet{rodu2018detecting} extended a kernel version of CCA (KCCA; \citealp{hardoon2004canonical}) to develop \emph{Dynamic Kernel CCA} (DKCCA), designed to detect time-varying connectivity between multivariate signals. 
\citet{gokcen2021disentangling} proposed \emph{Delayed Latents Across Groups} (DLAG), which extends GPFA to jointly model within- and across-region interactions using structured latent processes. Both methods come with own limitations. For instance, DLAG's reliance on a structured latent covariance can be limiting for systems with bi-directional coupling via low-dimensional oscillatory dynamics. 

In this work, we introduce a more flexible alternative by leaving the latent covariance unspecified, allowing for a fully data-driven estimate of cross-regional correlations. This yields a partial correlation graph representation of neural interactions. Given the high dimensionality of the possible partial correlations across time lags, we adopt a sparse estimation framework \citep{friedman2008sparse}, resulting in our method: {\it La}tent {\it Dyn}amic analysis via {\it S}parse banded graphs (LaDynS). {While we applied LaDynS to LFP recordings, it could also be used to analyze other slowly-varying multidimensional neural measurements made during repeated-trial cognitive tasks.}

In \cref{sec:2_methods}, after defining the model and proving that it produces a time series generalization of CCA, we give details of the fitting procedure and discuss inference using clustered contiguous significant association based on a de-biased estimate of the precision matrix \citep{jankova2015confidence}
together with control of false discovery rate  \citep{benjamini1995controlling}. We also use a state-space formulation to impose a local stationarity assumption \citep{ombao2022spectral}, pointing out how this can produce a time-varying latent Granger causality assessment.
Simulations in \cref{sec:simulation_from_SSM} show that our implementation of LaDynS is able to correctly identify the timing of interactions when applied to artificial data designed to be similar to those we analyzed, even in scenarios where existing methods exhibit low statistical power. The simulations support data-analytic results presented in \cref{sec:experimental_data}, which show significant cross-regional interaction at roughly $400$ ms after presentation of the stimulus (see \Cref{fig:result_Smith}).
This timing coincides with the delay in visual cortical response seen {after the onset of a visual stimulus} in mice \citep{chen2022population}, humans \citep{del2007brain,yang2019exploring}, and macaques \citep{super2001two,schmolesky1998signal}. The results reinforce the idea that PFC and V4 are involved, together, in working memory.
We add discussion in \cref{sec:4_conclusion}.

%% file: sections/2_methods.tex
\section{Methods}\label{sec:2_methods}

We begin by reviewing and reformulating probabilistic CCA (pCCA) in \cref{sec:pCCA_vectors}, which we then generalize to time series in \cref{sec:pCCAtimeseries} to 
yield a dynamic version of pCCA. We define the LaDynS model based on the loglikelihood function in \cref{def:ladyns}. We go over the choice of regularization parameters in \cref{sec:tuning_parameter}
and the fitting algorithm in \cref{sec:fit_algorithm}. We discuss statistical inference in \cref{sec:inference}.

\subsection{Probabilistic CCA for two random vectors} \label{sec:pCCA_vectors}

\begin{figure}
    \centering
    {\bf (a)}
    \raisebox{\dimexpr \topskip-\height}{
        \includegraphics[height = 0.07\textwidth]{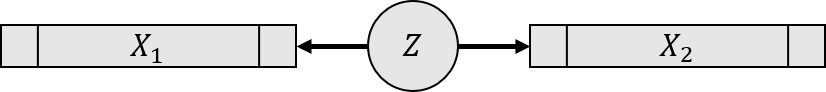}
    }
    
    \vspace{5mm}
    
    {\bf (b)}
    \raisebox{\dimexpr \topskip-\height}{
        \includegraphics[height = 0.07\textwidth]{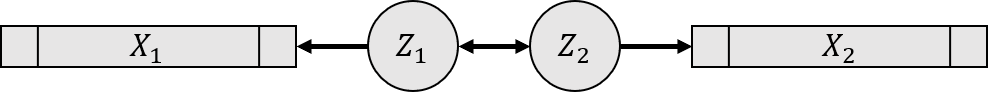}
    }
    
    \caption{{\bf Graphical representation of pCCA models.} \sl {\bf (a)} Model of \cite{bach2005probabilistic}, where $X_1$ and $X_2$ are random vectors and $Z$ is a random variable.  {\bf (b)} A variation on (a) that facilitates the extension to the case when $X_1$ and $X_2$ are multivariate time series and $(Z_1,Z_2)$ is a bivariate time series.}
    \label{fig:pCCA}
\end{figure}

Given two random vectors $X_1 \in \mathbb{R}^{d_1}$ and $X_2 \in \mathbb{R}^{d_2}$, canonical correlation analysis (CCA) \citep{hotelling1992relations} finds the sets of weights $w_1 \in \mathbb{R}^{d_1}$ and $w_2 \in \mathbb{R}^{d_2}$ that maximize Pearson's correlation 
between linear combinations $w_1^\top X_1$ and $w_2^\top X_2$. This can be rewritten as
\begin{equation}\label{eq:CCA}
    \sigma_{cc} = {\max}_{w_k, k=1,2 : w_{k}^\top \Sigma_{kk} w_k = 1} \abs*{w_1^\top \Sigma_{12} w_2},
\end{equation}
where 
$\Sigma_{kk}=\Var(X_k)$ is the covariance matrix of $X_k$, $k=1,2$, and $\Sigma_{12} = \Cov(X_1, X_2)$ the cross-covariance matrix between $X_1$ and $X_2$. The sample estimator $\hat{\sigma}_{cc}$ is obtained by replacing $\Sigma_{kk}$ and $\Sigma_{12}$ with their sample analogs $\bar{\Sigma}_{kk}$ and $\bar{\Sigma}_{12}$ respectively. The maximizing weights $\hat{w}_k$ and linear combinations $\hat{Z}_k = \hat{w}_k^\top X_k$ 
are referred to as the canonical weights and canonical variables, respectively.

Probabilistic CCA assumes that $X_1$ and $X_2$ are driven by a common one dimensional latent variable $Z$:
\begin{equation}\label{eq:pCCA_BJ}
\begin{aligned}
  X_k|Z & = \mu_k + Z \cdot \beta_k + \epsilon_k, ~ k = 1,2,  \\
  Z & \sim N(0,1),
\end{aligned}
\end{equation}
where $\mu_k \in \reals^{d_k}$ and $\beta_k \in \reals^{d_k}$ are mean vectors and factor loadings, respectively, and $\epsilon_k \distind \distMVN(0, \Phi_k)$ \citep{bach2005probabilistic}. 
\Cref{fig:pCCA}(a) depicts the dependence of $X_1$ and $X_2$ on $Z$. The parameters in \cref{eq:CCA,eq:pCCA_BJ} both yield the same estimate of $\sigma_{cc}$; see \cref{thm:equiv_of_BJ_to_CCA} for more details. 

Next, we introduce an alternative pCCA extension that assigns distinct latent variables for $X_1$ and $X_2$, as depicted in \Cref{fig:pCCA}(b). Specifically, we assume that
\begin{equation}\label{eq:pCCA_obs_proposed}
    X_k|Z_k = \mu_k + Z_k \cdot \beta_k + \epsilon_k
\end{equation}
where $\mu_k \in \reals^{d_k}$, $\beta_k \in \reals^{d_k}$ and $\epsilon_k \distind \distMVN(0, \Phi_k)$ are defined as in \cref{eq:pCCA_BJ}, and $(Z_1, Z_2)$ are bivariate normally distributed:
\begin{equation}\label{eq:pCCA_latent_proposed}
    \begin{pmatrix} Z_1 \\ Z_2 \end{pmatrix} \dist \distMVN \left(
    \begin{pmatrix} 0 \\ 0 \end{pmatrix}, 
    \begin{pmatrix} 
        1 & \sigma_{12} \\
        \sigma_{12} & 1 
    \end{pmatrix}\right).
\end{equation}
Like the original method, the alternative pCCA yields the same estimate of $ \sigma_{cc}$. We prove this equivalence in \cref{thm:equivalence_to_CCA}.
%

\subsection{Probabilistic CCA for two time series of random vectors \label{sec:pCCAtimeseries}}

\begin{figure}
    \centering
    {\bf (a)}
    \raisebox{\dimexpr \topskip-\height}{
        \includegraphics[width = 0.8\textwidth]{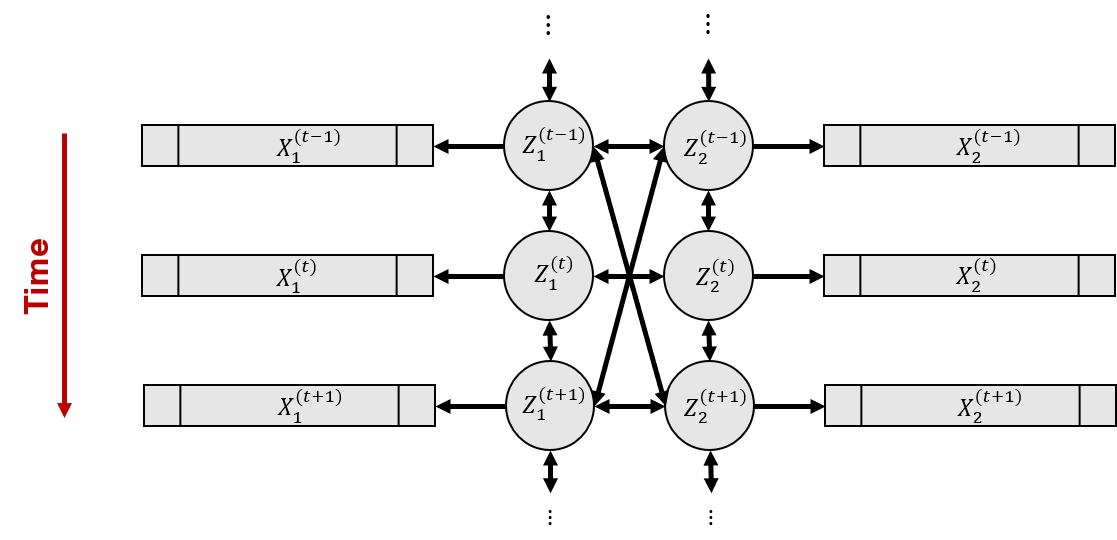}
    }
    
    \vspace{5mm}
    
    {\bf (b)}
    \raisebox{\dimexpr \topskip-\height}{
        \includegraphics[width = 0.8\textwidth]{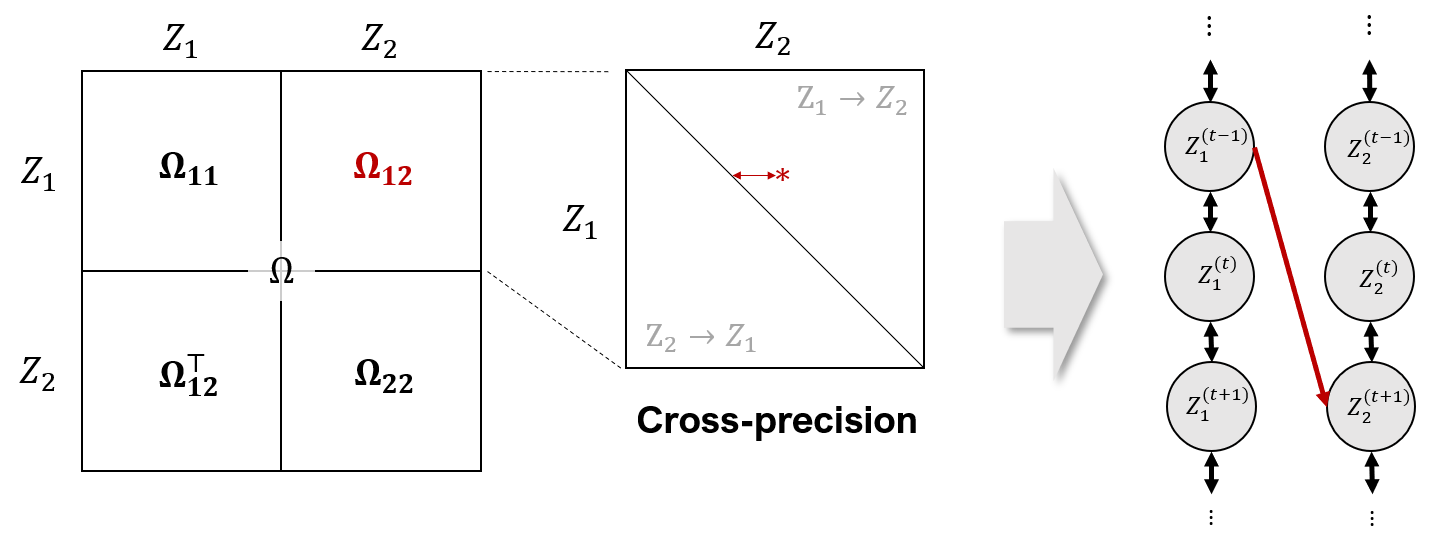}
    }

    \caption{\sl {\bf Extended pCCA model for two multivariate time series $X_1^{(t)}$ and $X_2^{(s)}$, $t, s=1, \ldots, T$.}
    {\bf(a)} Dynamic associations between vectors $X_1^{(t)}$ and $X_2^{(s)}$ are summarized by the dynamic associations between their associated 1-dimensional latent variables $Z_1^{(t)}$ and $Z_2^{(s)}$ represented by their  cross-precision matrix $\Omega_{12}$.
    {\bf(b)} When a significant cross-precision entry is identified, e.g., the red star in the expanded view of $\Omega_{12}$, its coordinates and distance from the diagonal indicate at what time in the experiment connectivity between two brain areas occurs, and at what lead or lag. 
    Here the red star is in the upper diagonal of $\Omega_{12}$, which means that, at this particular time, region 1 leads region 2, or $Z_1 \rightarrow Z_2$ in short (a non-zero entry in the lower diagonal would mean $Z_2 \rightarrow Z_1$). We represent this association by the red arrow on the right-most plot, with a lag of two units of time for illustration.
    }    
    \label{fig:pCCA_over_time}
\end{figure}

Suppose now that we are interested in the correlation dynamics between two times series of random vectors $X_1^{(t)} \in \reals^{d_1}$ and $X_2^{(t)} \in \reals^{d_2}$, $t=1,2,\dots,T$. For
each time $t$, we use \cref{eq:pCCA_obs_proposed} to model the dependence of $X_k^{(t)}$
on its associated latent variable $Z_k^{(t)}$:
\begin{equation}\label{eq:pCCA_obs_over_time}
    X_k^{(t)} | Z_k^{(t)} = \mu_k^{(t)} + \beta_k^{(t)} \cdot Z_k^{(t)} + \epsilon_k^{(t)}, ~~ k = 1, 2,
\end{equation}
where $\mu_k^{(t)}$, $\beta_k^{(t)}$ and  $\epsilon_k^{(t)} \distind \distMVN(0, \Phi_k^{(t)})$ are defined as in \cref{eq:pCCA_obs_proposed}.
%
Then for each $t$ we could define a parameter $\sigma_{12}^{(t)}$ as in \cref{eq:pCCA_latent_proposed}
to capture population-level association between $X_1^{(t)}$ and $X_2^{(t)}$ at $t$.
But because we are also interested in lagged associations between $X_1^{(t)}$ and $X_2^{(s)}$ for $s \neq t$, we replace the bivariate model (\ref{eq:pCCA_latent_proposed}) for $Z_1^{(t)}$ and $Z_2^{(t)}$ for a given $t$ by a global model for all $t=1, \ldots, T$:
\begin{equation} \label{eq:pCCA_latent_over_time}
    \left( \left(Z_1^{(t)}\right)_{t=1, \ldots, T}, \left(Z_2^{(t)}\right)_{t=1, \ldots, T }\right)^\top  \dist \distMVN(0, \Sigma), ~~ \diag(\Sigma) = \mathbf{1},
\end{equation}
where $\Sigma$ captures jointly all simultaneous and lagged associations within and between the two time series. 
\Cref{fig:pCCA_over_time}(a) illustrates the dependence structure of this model. 
%
%
We decompose ${\Sigma}$ and its inverse $\Omega$ as
\begin{equation}\label{eq:decomp_Sigma_Omega}
  \begingroup
  \def\arraystretch{1.3}
  {\Sigma} = \left(\begin{array}{c|c}
    {\Sigma}_{11} & {\Sigma}_{12} \\  \hline
    {\Sigma}_{12}^\top & {\Sigma}_{22}
  \end{array}\right)
  \endgroup
  \textand
  \begingroup
  \def\arraystretch{1.3}
  {\Omega} = \left(\begin{array}{c|c}
    {\Omega}_{11} & {\Omega}_{12} \\  \hline
    {\Omega}_{12}^\top & {\Omega}_{22}
  \end{array}\right)
  \endgroup
\end{equation}
to highlight the auto-correlations ${\Sigma}_{11}$ and ${\Sigma}_{22}$ within and cross-correlations $\Sigma_{12}$ between the time series, and denote by ${\Sigma}_{kl}^{(t,s)}$, $(t,s) \in [T]^2$, the elements of ${\Sigma}_{kl}$. Then ${\Sigma}_{12}^{(t,t)}$ for some fixed $t$ has the same interpretation as $\sigma_{12}$ in \cref{eq:pCCA_BJ}. 
Further, $-\Omega_{12}^{(t,s)}/\sqrt{\Omega_{11}^{(t,t)} \Omega_{22}^{(s,s)}}$ is the partial correlation between the two latent time series at times $t$ and $s$. Thus, when an element of $\Omega_{12}$ is non-null, depicted by the red star in the expanded display in \Cref{fig:pCCA_over_time}(b), its coordinates $(t,s)$ and distance $(t-s)$ from the diagonal indicate at what time in the trial a connectivity happens between two time series, and at what lead or lag, respectively. In our neuroscience application, they represent the timing of connections and direction of information flow between two brain regions. 
As is common with latent variable models, the signs of the latent covariance $\Sigma$ and therefore the latent precision $\Omega$ are not identifiable. 
This non-identifiability is benign for inference on the directionality in lead-lag connectivity, which is encoded in the support of $\Omega_{12}$, rather than by their signs. (See \Cref{fig:pCCA_over_time}(b).) The directionality remains interpretable even though the signs of the partial-correlations are unidentifiable.
Therefore in the following sections we focus on the estimation and inference for the magnitude of associations, rather than their sign.

As the equivalence between a non-distributional method (CCA) and its probabilistic representation (pCCA), there exists a similar connection between the multiset generalization of CCA introduced by \cite{kettenring1971canonical} and the dynamic pCCA model in \cref{eq:pCCA_obs_over_time,eq:pCCA_latent_over_time}. Multiset CCA applied to $2T$ random vectors $\{X_1^{(t)}, X_2^{(t)}:t=1,\dots,T\}$ finds weights $\{w_1^{(t)}, w_2^{(t)}:t=1,\dots,T\}$ that maximize a notion of correlation among linear combinations $\{w_1^{(t)\top} X_1^{(t)}, w_2^{(t)\top} X_2^{(t)}:t=1,\dots,T\}$.
%
In particular, the GENVAR extension minimizes the generalized variance of these linear combinations, defined as the determinant of their correlation matrix \citep{wilks1932certain}, which we refer to as the canonical correlation matrix:
\begin{eqnarray}\label{eq:mcca_genvar}
  \hspace{1cm} \hat w_1^{(1)}, \ldots, \hat w_2^{(T)} = \argmin_{w_1^{(1)}, \ldots, w_2^{(T)}} \det\left(\bar{\Var}\left[ \left( {w}_1^{(t)\top} X_1^{(t)}\right)_{t=1,\ldots,T}, \left( {w}_2^{(t)\top} X_2^{(t)}\right)_{t=1,\ldots,T} \right]\right)
\end{eqnarray}
where $\bar{\Var}$ denotes the sample variance-covariance matrix and the weights $w_k^{(t)}$ are scaled so that every diagonal entry of the matrix is $1$. 
In \cref{thm:equivalence_to_genvar} we show that the maximum likelihood estimators from the dynamic pCCA model recover the same canonical weights $\{w_1^{(1)}, \dots, w_2^{(T)}\}$ and the canonical correlation matrix $\hat\Sigma$ as the GENVAR formulation of multiset CCA. Under this equivalence the MLEs minimize
\begin{equation}\label{eq:nll_sigma_weight}
       \log\det(\Sigma) + \tr\left(\Sigma^{-1} \bar{\Sigma}\right), 
\end{equation}
an objective that motivates the Latent Dynamic Analysis via Sparse Banded Graphs (LaDynS) developed in the next section. The objective above does not involve any component of the observation model \cref{eq:pCCA_obs_over_time}; thus the estimators do not depend on a Gaussian assumption for $X_k^{(t)}\mid Z_k^{(t)}$.  
They only require approximate normality of the latent factors $Z_k^{(t)}$, which is often plausible since each $Z_k^{(t)}$ is a weighted sum of many coordinates of $X_k^{(t)}$.

\subsection{Latent Dynamic Analysis via Sparse Banded Graphs (LaDynS)}\label{sec:LaDynS}

\begin{figure}
\includegraphics[height=0.25\textwidth]{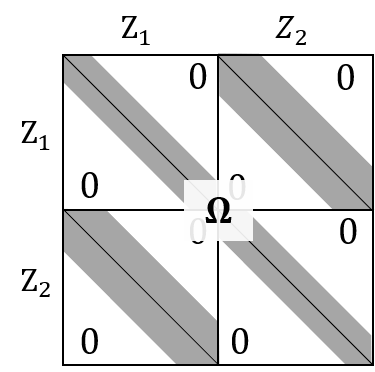}
\caption{\sl 
The elements of $\Omega_{kk}$, $k=1,2$, and $\Omega_{12}$ are set to zero outside of the gray bands of widths $(1+2d_\text{auto})$ and $(1+2d_\text{cross})$, respectively.}
\label{fig:forced_sparsity}
\end{figure}

Our goal is to estimate the association dynamics between two multivariate time series 
of length $T$ using the covariance matrix $\Sigma$ of their associated latent time series in~\cref{eq:pCCA_latent_over_time}. 
However, the prohibitive number of parameters in $\Sigma$ means its estimation is prone to errors, especially when $T$ is large. 
We reduce their number by regularizing $\Omega= \Sigma^{-1}$, rewriting $\log\det(\Sigma) =\log\det(\Omega^{-1}) = -\log\det(\Omega)$, and assuming that $\Omega$ has the banded structure depicted in \Cref{fig:forced_sparsity}.
\begin{definition}[LaDynS] \label{def:ladyns}
Given $N$ simultaneously recorded pairs of multivariate time series $\{X_{1[n]}, X_{2[n]}\}_{n=1,\dots,N}$, and a $2T \times 2T$ sparsity  matrix $\Lambda$ with element $\Lambda_{kl}^{(t,s)}$ regularizing $| \Omega_{kl}^{(t,s)} |$, $k, l = 1, 2$,
LaDynS finds weights $\left\{ \hat w_k^{(t)}, t=1,2,\dots,T, k=1,2\right\}$ and precision matrix $\hat \Omega$ that minimize the penalized negative log-likelihood:
\begin{equation}\label{eq:penalized_likelihood}
-\log\det(\Omega) + \tr(\Omega \bar{\Sigma}) + \|\Lambda \odot \Omega\|_1,
\end{equation}
where 
$\bar{\Sigma} = \bar{\Var}\left[w_1^{(1)\top} X_1^{(1)}, \dots, w_2^{(T)\top} X_2^{(T)}\right]$
satisfies $\diag(\bar\Sigma) = \mathbf{1}$,
$\odot$ denotes the Hadamard product operator such that $(A \odot B)_{ij} = A_{ij} \times B_{ij}$, $\|A\|_1 = \sum_{i,j} |A_{ij}|$,
and
\begin{equation*}
  \Lambda_{kl}^{(t,s)} = \begin{cases}
      \lambda_\text{cross}, & k \neq l ~~\text{and}~~ 0< \abs{t-s} \leq d_\text{cross},   \\
      \lambda_\text{auto}, &  k = l ~~\text{and}~~ 0< \abs{t-s} \leq d_\text{auto}, \\
    \lambda_\text{diag}, & t = s, \\
    \infty, & \text{otherwise},
  \end{cases}
\end{equation*}
which constrains auto-precision and cross-precision elements within a specified range.
\end{definition}
%
In our neuroscience application, in particular, it is reasonable to assume that lead-lag relationships 
occur with delay less than temporal bandwidth $d_\text{cross}$, 
which can be determined by the maximal transmission time in synaptic connections between two brain regions under study.
We thus set $\Lambda_{12}^{(t,s)} = \infty$ when $\abs{t-s} > d_\text{cross}$ to force the corresponding cross-precision elements to zero and thus 
impose a banded structure on $\Omega_{12}$. 
We apply sparsity constraint $\Lambda_{12}^{(t,s)} = \lambda_\text{cross} > 0 $ 
on the remaining off-diagonals of $\Omega_{12}$ to focus our discovery 
of sparse dominant associations and reduce the effective parameter size.
%
We proceed similarly with the auto-precision matrices $\Omega_{11}$ and $\Omega_{22}$, using penalty $\lambda_\text{auto}$ and temporal bandwidth $d_\text{auto}$. Unless domain knowledge is available, we recommend that $d_\text{auto}$ 
be set to the largest significant auto-correlation across all observed time series $X_{k,i}^{(t)}$, $k=1,2$, $i=1, \ldots, N$, 
and impose no further sparsity ($\lambda_\text{auto} = 0$) unless there is reason to expect it.
%

%
%
Notice that, for simplicity, we grouped the elements of $\Lambda$ into diagonal and off-diagonal elements and assigned the same penalties, $\lambda_\text{cross}$, $\lambda_\text{auto}=0$ and $\lambda_\text{diag}$, within each group. 

{
Again, LaDynS estimates the magnitude of the associations, but their signs remain non-identifiable. Accordingly, our results in \cref{sec:results} report and display the absolute values of the estimated variance and precision elements.
}

\subsubsection{Choosing regularization parameters}\label{sec:tuning_parameter}

In graphical LASSO (gLASSO) problems, where the aim is to recover correct partial correlation graphs, penalties are often chosen 
to minimize the predictive risk \citep{shao1993linear,zou2007degrees,tibshirani2012degrees}.
Our aim is different: only the partial cross-precision matrix $\Omega_{12}$ is of substantive interest, and because minimizing the predictive risk does not select models consistently \citep{shao1993linear, zhu2018sparse} and may 
thus fail to retrieve non-zero elements of $\Omega_{12}$, we 
choose instead a value of $\lambda_\text{cross}$ that controls the number of 
false cross-precision discoveries. 
We proceed by permuting the observed time series in one brain region to create a synthetic dataset that contains no cross-region correlation,
then applying LaDynS to that data for a range of values of $\lambda_\text{cross}$ and recording the resulting number of significant partial correlation estimates, which are necessarily spurious. We use the smallest $\lambda_\text{cross}$ that yields 
fewer false discoveries than a chosen threshold. 
We expect this regularization to make similarly few false discoveries on experimental data. 

%
Finally, if $\hat\Sigma$ cannot be inverted, as is the case for the band-pass filtered experimental data we analyze in \cref{sec:results}, we penalize its diagonal by $\lambda_\text{diag} > 0$. We explain the specific calibration 
we used for the analyzed datasets and study the properties in \cref{sec:simulation_result}.
%

\subsubsection{Fitting LaDynS using coordinate descent} \label{sec:fit_algorithm}

\cref{eq:penalized_likelihood} is not a convex function of the weights and precision elements (although it is not impossible that it may be for some particular $\Sigma$) and its convex relaxation is unknown, so it is difficult to find its global minimum. The coordinate descent algorithm described below finds a minimum, possibly local since it may be sensitive to the choice of initial parameter values. However, the simulation in \cref{app:sensitivity_analysis} suggests that the algorithm's sensitivity to initial values has little impact on its ability 
to recover the correct connectivity.

Assuming that all canonical weights $w_k^{(t)}$ are fixed, \cref{eq:penalized_likelihood} reduces to the gLASSO problem:
\begin{equation}\label{eq:GLasso}
     \argmin_{\Omega} -\log\det(\Omega) + \tr(\Omega \bar{\Sigma}) + \|\Lambda \odot \Omega\|_1,
\end{equation}
which we can solve efficiently using a number of existing algorithms; here we use the P-gLASSO algorithm of \cite{mazumder2012graphical}.
Then assuming that all parameters are fixed but a single weight $w_k^{(t)}$, \cref{eq:penalized_likelihood} can be re-arranged as the linear problem:
\begin{equation}\label{eq:LaDynS_w}
%
\argmin_{w_k^{(t)}} {\sum}_{(l, s) \neq (k, t)} w_k^{(t)\top} \bar{\Cov}\left[X_k^{(t)}, X_{l}^{(s)}\right] w_{l}^{(s)} \Omega_{kl}^{(t,s)}  ~~\text{s.t.}~~ w_k^{(t)\top} \bar{\Var}[X_k^{(t)}] w_k^{(t)} = 1,
\end{equation}
for which an analytical solution is available.
That is, our algorithm alternates between updating $\Omega$ and the weights $w_k^{(t)}$ until the objective function in \cref{eq:penalized_likelihood} converges.
Its computational cost is inexpensive: a single iteration on our cluster server (with $11$ Intel(R) Xeon(R) CPU \@ $2.90$ GHz processors) took in average less than $0.8$ seconds, applied to the experimental data in \cref{sec:experimental_data}. A single fit on the same data took $47$ iterations for around $33.57$ seconds until the objective function converged at threshold $0.001$.
\cref{alg:LaDynS_simplified} presents a high level pseudocode sketch of the coordinate descent algorithm. See \cref{app:LaDynS_algorithm} for details and Python package \texttt{ladyns} on \url{github.com/HeejongBong/ladyns}.

\begin{algorithm}[h]
\caption{Coordinate descent algorithm to fit LaDynS}
\begin{algorithmic}[1]
    \INPUT \hfill\\
    $\{X_k: k = 1, \dots, K\}$: input data \\
    $\Lambda \in [0,\infty]^{KT \times KT}$: sparsity penalty matrix \\
    $\text{iter}_\text{max} \in \nats_{+}$: maximum iteration \\
    $\text{ths} \in \reals_{+}$: threshold for convergence
    \\\hfill\\
    \OUTPUT $\Omega$ and $w_k^{(t)}$'s which solve \cref{eq:penalized_likelihood} w.r.t. \cref{fig:forced_sparsity}
    \\\hfill\\
    \INIT
    \STATE Initialize $w_k^{(t)}$ so that $w_k^{(t)\top} \bar{\Var}[X_k^{(t)}] w_k^{(t)} = 1$ for all $t \in [T]$ and $k = 1, 2$. e.g.,
    \begin{equation*}
        w_k^{(t)} \leftarrow  \mathbf{1} / \sqrt{\mathbf{1}^\top \bar{\Var}[X_k^{(t)}] \mathbf{1}} 
        ~~\text{and}~~
        \bar\Sigma \leftarrow \bar{\Var}[w_1^{(1)\top} X_1^{(1)}, \dots, w_2^{(T)\top} X_2^{(T)}].
    \end{equation*}
    \ITER
    \FOR{$\text{iter}$ in 1:$\text{iter}_\text{max}$}
        \STATE $\Omega \leftarrow \argmin_{\Omega'} -\log\det(\Omega') + \tr(\Omega' \bar{\Sigma}) + \|\Lambda \odot \Omega'\|_1$.
        \\\hfill\\
        \FOR{$k$ in 1:2 and $t$ in 1:$T$}
            %
            \STATE $w_k^{(t)} \leftarrow \argmin_{w} {\sum}_{(l, s) \neq (k, t)} w^{\top} \bar{\Cov}\left[X_k^{(t)}, X_{l}^{(s)}\right] w_{l}^{(s)} \Omega_{kl}^{(t,s)}  ~~\text{s.t.}~~ w^{\top} \bar{\Var}[X_k^{(t)}] w = 1,$
        \ENDFOR
        \\\hfill\\
        \STATE $\bar \Sigma \leftarrow \bar{\Var}\left[w_1^{(1)\top} X_1^{(1)}, \dots, w_2^{(T)\top} X_2^{(T)}\right].$
        \\\hfill\\
        \IF{$\max(|\Sigma - \Sigma_{\text{last}}|) < \text{ths}$}
            \STATE \textbf{break}
        \ENDIF
    \ENDFOR
\end{algorithmic}
\label{alg:LaDynS_simplified}
\end{algorithm}

\subsection{Inference for associations between two vector time series} \label{sec:inference}

Let $\hat{\Omega}$ and $\hat{w}_k^{(t)}$, $t=1,\ldots,T$, $k=1,2$, be the LaDynS estimates of canonical precision matrix and canonical weights, and $\bar{\Sigma} = \bar{\Var}\left[\hat{w}_1^{(1)\top} X_1^{(1)}, \dots, \hat{w}_2^{(T)\top} X_2^{(T)}\right]$ be the empirical covariance of the estimated latent variables, defined in \cref{eq:penalized_likelihood}.
%
Note that $\hat{\Omega} \neq \bar{\Sigma}^{-1}$ since $\hat\Omega$ is constrained to be sparse. Based on these estimates, 
we want to identify the non-zero partial cross-correlations in $\Omega_{12}$, 
which correspond to the epochs of association between the two time series.



Formal inference methods 
for $\Omega$ based on its LaDynS estimate (\cref{eq:penalized_likelihood}) are not available, 
but because LaDynS reduces to graphical LASSO (gLASSO) when the weights $w_k^{(t)}$ in \cref{eq:GLasso} are fixed, 
we co-opt gLASSO inference methods.
%
Specifically, \cite{jankova2015confidence} suggested de-sparsifying the gLASSO estimate $\hat \Omega$ according to
\begin{equation}\label{eq:desparsified_precision}
    \tilde{\Omega} = 2\hat{\Omega} - \hat{\Omega}(\bar{\Sigma} + \lambda_\text{diag} I_T) \hat{\Omega},
\end{equation}
and proved that, under mild assumptions and as $n \rightarrow \infty$, each entry of $\tilde{\Omega}$ satisfies the Central Limit Theorem with center the true precision $\Omega$: 
\begin{equation}
\label{eq:CLT_on_dspr_LaDynS}
  \forall (t,s), ~~    \frac{\left(\tilde{\Omega}_{12}^{(t,s)} - \Omega_{12}^{(t,s)} \right)}{\sqrt{\Var[\tilde{\Omega}_{12}^{(t,s)}]}} \overset{d}{\rightarrow} \distNorm(0,1). 
\end{equation}
The assumptions include independence across experimental trials, irrepresentability condition and bounded eigenvalues of $\Omega$ (see Assumptions (A1) and (A2) in \citet{jankova2015confidence}). While some of these conditions may be strong and difficult to verify in practice, they are commonly imposed in the literature on sparse graph estimation with $\ell_1$ regularization. In \cref{app:simulation_from_model}, we examine the asymptotic normality of the de-sparsified LaDynS estimate using simulation.
%
We applied this result to the de-sparsified LaDynS estimate of $\Omega$, even though we do not quite have a gLASSO setup, and we verified
by simulation that its elements are indeed approximately normal in \cref{sec:simulation_result}.
\cite{jankova2015confidence} also proposed an estimator of $\Var[\tilde{\Omega}_{12}^{(t,s)}]$
that is likely to be downward biased in our framework since estimating the canonical 
weights $w_k^{(t)}$ induces extra randomness.
Instead, we use the bootstrap estimate $\hat{\Var}[\tilde{\Omega}_{12}^{(t,s)}]$ described at the end of this section.
We then 
rely on \cref{eq:CLT_on_dspr_LaDynS} to obtain p-values
\begin{equation}\label{eq:p_prec}
    p^{(t,s)} = 2-2 
    \Phi\left( | \tilde{\Omega}_{12}^{(t,s)} | \left/ \sqrt{\hat\Var[\tilde{\Omega}_{12}^{(t,s)}]}\right) \right.
\end{equation}
to test $H^{(t,s)}_0: \Omega_{12}^{(t,s)} = 0$, for each $(t,s) \in [T]^2$ within $d_\text{cross}$ of the diagonal of $\Omega_{12}$.

\smallskip

\noindent{\it Permutation bootstrap estimate of $~\Var[\tilde{\Omega}_{12}^{(t,s)}]$: }
A permutation bootstrap sample $\{X^*_{1[n]}, X^*_{2[n]}\}_{n=1,\dots,N}$ is generated 
by permuting separately $\{X_{1[n]}\}_{n=1,\dots,N}$ and $\{X_{2[n]}\}_{n=1,\dots,N}$.
Hence, applying 
LaDynS to $\{X^*_{1[n]}, X^*_{2[n]}\}_{n=1,\dots,N}$ yields estimates of canonical precision matrix $\hat{\Omega}^*$, canonical weights $\hat{w}_k^*(t)$s, empirical covariance of the estimated latent variables $\bar{\Sigma}^{*} = \bar{\Var} \left(\hat{w}_1^{*(1)\top} X_1^{*(1)}, \dots, \hat{w}_2^{*(T)\top} X_2^{*(T)} \right)$, and de-sparsified precision matrix estimate $\tilde{\Omega}^*$ (\cref{eq:desparsified_precision}) under the global null hypothesis of no correlated activity.
Repeating the bootstrap simulation $B$ times produces $B$ bootstrap values $\hat{\Omega}^{b}$, $\hat{w}_k^{b(t)}$, $\bar{\Sigma}^{b}$, and $\tilde{\Omega}^b$, $b=1, \ldots, B$.
We then estimate $\Var[\tilde{\Omega}_{12}^{(t,s)}]$ with $\hat\Var[\tilde{\Omega}_{12}^{(t,s)}]$, 
the sample standard error of $\{\tilde{\Omega}_{12}^{b,(t,s)}\}_{b=1,2,\dots,B}$.
Notice that $\hat{\Var}[\tilde{\Omega}_{12}^{(t,s)}]$ is obtained under the global null $H^{(t,s)}_0: \Omega_{12}^{(t,s)} = 0$ simultaneously for all $(t,s)$, because it is not trivial to simulate bootstrap data that satisfy a specific $H^{(t,s)}_0$ without 
assuming that all other elements of $\Omega_{12}$ are also null.
We garnered from simulations that $\hat{\Var}[\tilde{\Omega}_{12}^{(t,s)}]$ is thus likely to slightly underestimate $\Var[\tilde{\Omega}_{12}^{(t,s)}]$, which makes for slightly sensitive p-values. 

\smallskip

\noindent{\it Control of false discoveries: }
Because we perform tests for many entries of $\Omega_{12}$, 
we cap the false discovery rate
\begin{equation} \label{eq:FDR_FDP}
\text{FDR} = \Exp\left[ \text{FDP} \right], ~~ \text{where} ~~ \text{FDP}=\frac{\#\{\text{falsely discovered entries}\}}{\#\{\text{discovered entries}\} \vee 1}
\end{equation}
below a pre-specified level $\alpha_\text{BH}$ using the procedure of \cite{benjamini1995controlling} (BH). 
To proceed, let ${p}^{[1]} \leq \dots \leq {p}^{[n_\text{roi}]}$ denote the ordered permutation bootstrap 
p-values ${p}^{(t,s)}$ that correspond to $n_\text{roi}$ cross-precision elements in the region of interest. 
Then, we find the maximum $k_\text{BH}$ satisfying ${p}^{[k_\text{BH}]} \leq \frac{k_\text{BH}}{n_\text{roi}} \alpha_\text{BH}$ and reject $H_0^{(t,s)}$ with ${p}^{(t,s)}$ smaller than $\frac{k_\text{BH}}{n_\text{roi}} \alpha_\text{BH}$. The FDR guarantee is established by \citet{benjamini1995controlling} as long as the ${p}^{(t,s)}$'s are independent and valid p-values. 


\smallskip

\noindent{\it Cluster-wise inference by excursion test: }
As a further safeguard against falsely detecting correlated activity between brain areas, we obtain p-values for each identified connectivity epoch using the excursion test of \cite{ventura2005statistical}, as follows. 
%
%
After computing p values $p^{(t,s)}$ for each precision entry (\cref{eq:p_prec}), we define a cluster as any contiguous set of entries whose p-values fall below $\frac{k_{\mathrm{BH}}}{n_{\mathrm{roi}}}\alpha_{\mathrm{BH}}$. No additional cluster forming threshold or minimum cluster size is imposed.
For each cluster $k$ identified by the BH procedure, we calculate the test statistic:
\begin{equation}
    T_k :=  - 2 \sum_{(t,s) \in \text{cluster $k$}} \log p^{(t,s)},
    \label{eq:exc}
\end{equation}
which is reminiscent of Fisher's method for testing the global significance of multiple hypotheses \citep{fisher1925statistical}.
Large values of $T_k$ provide evidence against cross-area connectivity in cluster $k$,
so we calculate the corresponding p-value as $\int_{T_k}^\infty f_{T_{\max}}(t) ~dt$, 
where $f_{T_{\max}}$ is the null distribution of $T_{\max} := \max_j T_j$ under the global null hypothesis
of no connectivity anywhere. We use $f_{T_{\max}}$ rather than the respective null distributions of each $T_k$
to control the family-wise type I error rather than the type I error for each cluster.
We approximate $f_{T_{\max}}$ by the previous permutation bootstrap: for each permuted dataset $b=1, \ldots, B$, we estimate the cross-precision matrix and corresponding p-values, identify
all clusters of p-values below $\frac{k_\text{BH}}{n_\text{roi}} \alpha_\text{BH}$, calculate the corresponding test statistics in \cref{eq:exc}, 
and let $T_{\max}^b$ be their maximum. The $B$ values $T_{\max}^b$ are samples from $f_{T_{\max}}$, which we use to approximate the p-value for 
cluster $k$ by the sampling
proportion:
\begin{equation*}
    \frac{1}{B} \sum_{b=1}^B \mathbb{I}(T_{\max}^b \geq T_k).
\end{equation*}


\subsection{Locally Stationary State-space Model and Local Granger Causality}
\label{sec:local_granger_causality}

Our model in \cref{eq:pCCA_obs_over_time,eq:pCCA_latent_over_time} can be formulated as a state-space model
by rewriting the joint multivariate Gaussian model for the latent vectors in \cref{eq:pCCA_latent_over_time} 
as the set of all
conditional distributions
\begin{equation} \label{eq:AR_latent_model}
\begin{aligned}
    Z_1^{(t)} & = \sum_{s=1}^{d_\mathrm{auto}} \alpha_{11,s}^{(t)} Z_1^{(t-s)}
    + \sum_{s=1}^{d_\mathrm{cross}} \alpha_{12,s}^{(t)} Z_2^{(t-s)}
    + \eta_1^{(t)}, \\
    Z_2^{(t)} & = \sum_{s=1}^{d_\mathrm{auto}} \alpha_{22,s}^{(t)} Z_2^{(t-s)}
    + \sum_{s=1}^{d_\mathrm{cross}} \alpha_{21,s}^{(t)} Z_1^{(t-s)}
    + \eta_2^{(t)}, \\
\end{aligned}
\end{equation}
where $d_\mathrm{auto}$ and $d_\mathrm{cross}$ are the maximal time delays in within-region 
and cross-region connections, $\eta_k^{(t)}$, $k=1, 2$, are independent $N(0, \phi_k^{(t)})$ random variables, 
and the $\alpha_{kl,s}^{(t)}$'s are vector auto-regressive coefficient parameters 
for the auto-correlation within region if $k=l$, $k=1,2$, 
and cross-correlation between regions if $k \neq l$ with time lag $s$. 

This state-space formulation is convenient to impose local stationarity on the latent time series,
which we do 
by fitting stationary state-space models in moving windows of time. 
Local stationarity is justified
because the functional connectivity within and between brain regions changes relatively slowly over time.
The state-space formulation is also convenient to calculate the Granger causality between regions:
$Z_2$ is said to Granger-cause $Z_1$ at time $t$ if some $\alpha_{12,s}^{(t)}$ are non-zero \citep{ombao2022spectral} (and 
conversely if some $\alpha_{21,s}^{(t)}$ are non-zero). The coefficient of partial determination (partial $R^2$) between $\left(Z_2^{(t-d_\mathrm{cross})}, \dots, Z_2^{(t-1)}\right)$ and $Z_1^{(t)}$, conditional on $Z_1^{(t-d_\mathrm{auto})}, \dots, Z_1^{(t-1)}$, 
calculated as
\begin{equation}
    R^2_{2 \rightarrow 1}(t) = 1 - \frac{\Var[\text{residual of Regression 1}]} {\Var[\text{residual of Regression 2}]},
\end{equation}
where
\begin{equation}
\begin{aligned}
    \text{Regression 1}: Z_1^{(t)}
    \sim & Z_1^{(t-d_\mathrm{auto})} + \dots + Z_1^{(t-1)} 
    + Z_2^{(t-d_\mathrm{cross})} + \dots + Z_2^{(t-1)}, \\
    \text{Regression 2}: Z_1^{(t)} 
    \sim & Z_1^{(t-d_\mathrm{auto})} + \dots + Z_1^{(t-1)},
\end{aligned}
\end{equation}
may therefore be considered a test statistic for local Granger causality at time $t$. 
To allow a physiologically meaningful minimum connection time $\tau_1$ from brain regions 2 to 1, we can also 
replace the second regression by
\begin{equation*}
\begin{aligned}
    \text{Regression 2}: Z_1^{(t)} 
    \sim & Z_1^{(t-d_\mathrm{auto})} + \dots + Z_1^{(t-1)} + Z_2^{(t-\tau_1+1)} + \dots + Z_2^{(t-1)} \\
    + & \mathbf{1}\{\tau_2 < d_\mathrm{cross}\} \left(Z_2^{(t-d_\mathrm{cross})} + \dots + Z_2^{(t-\tau_2-1)} \right)
    , \\
\end{aligned}
\end{equation*}
where $\tau_2$ is the maximum connection time from brain 
regions 2 to 1, $\tau_1 \le \tau_2 \le d_\mathrm{cross}$.
We set $\tau_2= d_\mathrm{cross}$ by default, unless there is reason to consider shorter 
connection times.
A plug-in estimator of $R^2_{2 \rightarrow 1}(t)$ is easily obtained from the estimated covariance matrix of 
$
    \left(Z_1^{(t)}, \dots, Z_1^{(t-d_\mathrm{auto})}, Z_2^{(t-1)}, \dots, Z_2^{(t-d_\mathrm{cross})}\right),
$
without actually running the regressions \citep{anderson1994model}. 

Autocorrelations in the latent time series can inflate $R^2$ values.
We therefore test the statistical significance of $R^2_{2 \rightarrow 1}(t)$ (or $R^2_{1 \rightarrow 2}(t)$) by
comparing its
observed value to its null distribution, obtained by repeatedly permuting the trials in one region and calculating 
$R^2_{2 \rightarrow 1}(t)$ in the permuted data. We used 
2000 permutations in \cref{sec:experimental_data}. The permuted data satisfy the null hypothesis of no cross-region connection
and exhibit the same autocorrelation structure as the original latent time series. 
 


%% file: sections/3_results.tex
\section{Results} \label{sec:results}

We have introduced LaDynS to estimate the dynamic connectivity between two or more multivariate time series, and we proposed inference procedures to identify when connectivity is statistically significant.
We apply LaDynS to experimental data in \cref{sec:experimental_data}, but first we present the theoretical results on the equivalence between generative pCCA models and model-free CCA methods. In particular, \cref{thm:equivalence_to_genvar} establishes an equivalence between the GENVAR version of multi-set CCA \citep{kettenring1971canonical} and maximum likelihood applied to our dynamic pCCA.
Next we examine its performance on simulated data that have properties similar
to the experimental data. 
The data are simulated from the shared oscillatory 
driver model described in \cref{sec:simulation_from_SSM}, 
which is unrelated to
the LaDynS model. \Cref{app:simulation_from_model} contains results based on simulated data that are consistent with the LaDynS model.
In \cref{sec:comparison}, we compare the performance of LaDynS to other existing methods.
The reproducible code scripts for the simulations and experimental data analyses are provided at \url{github.com/HeejongBong/ladyns}.

\subsection{Equivalence between dynamic pCCA model and multiset CCA} \label{sec:theoretical_results}

We first revisit the connection between classical CCA and the probabilistic CCA formulation of \citet{bach2005probabilistic}.  
Let $\{(X_{1[n]}, X_{2[n]})\}_{n=1}^N$ be independent observations drawn from the joint distribution in \cref{eq:pCCA_BJ}, and denote the maximum likelihood estimates by $(\hat\beta_1, \hat\beta_2)$.  
The result below restates Theorem 2 of \citet{bach2005probabilistic}.
\begin{theorem}[\citealp{bach2005probabilistic}, Theorem 2] \label{thm:equiv_of_BJ_to_CCA}
The maximum likelihood estimators (MLEs) 
$(\hat{\beta}_1, \hat{\beta}_2)$ in \cref{eq:pCCA_BJ} based on $N$ observed vector pairs $\left\{X_{1[n]}, X_{2[n]} \right\}_{n=1,2,\dots,N}$ are equivalent to the CCA solution $(\hat{w}_1, \hat{w}_2, \hat{\sigma}_{cc})$ in \cref{eq:CCA} according to:
\begin{equation}
    \hat{\beta}_k = \bar{\Sigma}_{kk} \hat{w}_k m_k, ~ \text{where} ~~ m_1 m_2 = \hat{\sigma}_{cc} ~~\text{and}~~ |m_k| \leq 1, ~ k=1,2.
\end{equation}
\end{theorem}
\cref{thm:equiv_of_BJ_to_CCA} proves that the original CCA setting and the generative pCCA model both yield the same estimate of $\sigma_{cc}$. 

We now state an equivalence similar to \cref{thm:equiv_of_BJ_to_CCA} between the original CCA and the alternative pCCA model.
\begin{theorem}\label{thm:equivalence_to_CCA}
The MLEs $(\hat{\beta}_1, \hat{\beta}_2, \hat{\sigma}_{12})$ in \cref{eq:pCCA_obs_proposed,eq:pCCA_latent_proposed} based on $N$ observed vector pairs $\left\{X_{1,[n]}, X_{2,[n]} \right\}_{n=1,2,\dots,N}$ are equivalent to the CCA solution $(\hat{w}_1, \hat{w}_2, \hat{\sigma}_{cc})$ according to:
\begin{equation}
\hat{\beta}_k = \bar{\Sigma}_{kk} \hat{w}_k m_k, ~ \text{where} ~~ m_1 m_2 \hat{\sigma}_{12} = \hat{\sigma}_{cc} ~~\text{and}~~ |m_k| \leq 1, ~ k=1,2.
\end{equation}
\end{theorem}
In practice we take $m_1 = m_2 = 1$ out of all possible solutions, because then
$Z_k | X_k = \hat{w}^\top_k X_k$ is the canonical variable almost surely, 
and $\sigma_{12} = \Cov[Z_1, Z_2]$ equals the canonical correlation $\sigma_{cc}$.
This means that 
$\sigma_{12}$ is an interpretable parameter, 
and one for which inference is simpler than for the canonical correlation in the other model, 
because, in \cref{eq:pCCA_BJ}, $\hat {\sigma}_{cc}$ is an indirect function of the maximum-likelihood parameter estimates (see \cref{thm:equiv_of_BJ_to_CCA}).
The interpretability property also persists when we extend model (\ref{eq:pCCA_latent_proposed}) to capture lagged association between two vector time series (see model (\ref{eq:pCCA_latent_over_time})).
Finally, the choice $m_1 = m_2 = 1$ implies that 
the MLEs $(\hat{\beta}_1, \hat{\beta}_2, \hat{\sigma}_{12})$ do not depend on the Gaussian assumption in \cref{eq:pCCA_obs_proposed}, 
an assumption that is questionable if, for
example, the $X$'s
are positive variables like LFP power envelopes or discrete variables like spike counts.
\cref{thm:equivalence_to_CCA} is recovered as a special case of \cref{thm:equivalence_to_genvar} below, which extends the result to multiset CCA for vector time series.

We now derive a similar connection between the multiset generalization of CCA introduced by \cite{kettenring1971canonical} and the dynamic pCCA model in \cref{eq:pCCA_obs_over_time,eq:pCCA_latent_over_time}. The proof is provided in \cref{app:pf_thm_equiv_to_genvar}.  

\begin{theorem} \label{thm:equivalence_to_genvar}
Suppose that $\hat \beta_k^{(t)}, ~k=1,2, ~t=1, \ldots, T$, and $\hat{\Sigma}$ are the MLE in \cref{eq:pCCA_obs_over_time,eq:pCCA_latent_over_time}
based on $N$ observed pairs of vector time series $\left\{X_{1[n]}^{(t)}, X_{2[n]}^{(t)}:~ t=1,\ldots,T \right\}$, $n=1,\dots,N$.
Then, they are equivalent to GENVAR multiset CCA solution according to:
\begin{equation}\label{eq:genvar_solution}
    \hat{\beta}_k^{(t)} = \bar{\Var} [ X_k^{(t)} ] \hat{w}_k^{(t)} m_k^{(t)} \textand \hat{\Sigma}_{kl}^{(t,s)} =
    \begin{cases}
        1, & k=l \textand t=s, \\
        \bar{\Var}\left[ \hat{Z}_k^{(t)}, \hat{Z}_l^{(s)} \right] , & \text{elsewhere,}
    \end{cases}
\end{equation}
where the canonical variable is
\begin{equation} \label{eq:canonical_variable}
    \hat{Z}_k^{(t)} = \hat{w}_k^{(t)\top} X_k^{(t)} = \frac{1}{m_k^{(t)}} \hat{\beta}_k^{(t)\top}\bar{\Var}^{-1}[X_k^{(t)}] X_k^{(t)},
\end{equation}
and $\abs{m_k^{(t)}} \leq 1$ for $k = 1, 2$ and $t \in [T]$. Furthermore, if all $m_k^{(t)} = 1$, then the MLE minimizes
\begin{equation}
       \log\det(\Sigma) + \tr\left(\Sigma^{-1} \bar{\Sigma}\right), 
\end{equation}
where $\bar{\Sigma} = \bar{\Var}\left[ \left( {\beta}_1^{(t)\top} \bar{\Var}^{-1}[X_1^{(t)}] X_1^{(t)}\right)_{t=1,\ldots,T}, \left({\beta}_2^{(t)\top} \bar{\Var}^{-1}[X_2^{(t)}] X_2^{(t)}\right)_{t=1,\ldots,T} \right]$, with $\beta_1$ and $\beta_2$ scaled such that $\diag{\bar\Sigma} = \mathbf{1}$.
\end{theorem}

\subsection{LaDynS performance on simulated data from a shared oscillatory driver model with time delay}
\label{sec:simulation_from_SSM}

We used a probabilistic model of the inter-areal coherence in local field potentials to generate LFP time series with a dynamic lead-lag relationship between two brain areas.
The shared oscillatory driver model with time delay in \citet{ombao2022spectral} assumes that the coherence between two univariate LFP time series $L_1$ and $L_2$ is driven by a univariate latent oscillation $L_0$ at frequency $f_0$:
\begin{equation} \label{eq:lagged_mixture_model}
\begin{aligned}
    L_k^{(t)} = \beta_k \cdot L_0^{(t-\tau_k)} + \eta_k^{(t)}, ~~ k=1,2,
\end{aligned}
\end{equation}
where $\eta_k^{(t)}$ is regional baseline noise, and 
$\tau_k$ is the lead-lag from $L_0$ to $L_k$.
This model generalizes the Synaptic-Source-Mixing (SSM) model of \citet{schneider2021mechanism}, which was shown to be capable of modeling the dynamic cross-regional coherence observed in experimental LFP data between frontal and parietal cortices, as
well as between LGN and the visual cortex.  
 
For our simulation, we extended \cref{eq:lagged_mixture_model} to generate two sets of multi-dimensional LFP 
time series $L_k$ of dimension $d_k$, $k=1,2$, and 
we used three latent oscillations $L_{0,j}$, 
$j=1,2,3$, to allow non-stationary lead-lag relationships 
between the two brain areas. The resulting model was a special case of the Latent Dynamic Factor Analysis model in \citet{bong2020latent}:
\begin{equation} \label{eq:nonstationary_lagged_mixture_model}
    L_k^{(t)} = \sum_{j=1}^3 \beta_{kj} \cdot L_{0,j}^{(t-\tau_{kj})} + \eta_k^{(t)}, ~~ k=1,2,
\end{equation}
where the factor loadings $\beta_{kj} \in \reals^{d_k}$ were vectors of dimension $d_k=25$.
We virtually arranged the $25$ components of $L_1^{(t)}$ 
and $L_2^{(t)}$
on $5 \times 5$ regular two-dimensional arrays
so we could create simulated data exhibiting spatial correlations 
similar to those in the experimental data in \cref{sec:experimental_data}.
To do that we let $\eta_k^{(t)}$ be spatially correlated Gaussian noise whose temporal power spectrum scales as $1/f^\alpha$ for $\alpha > 1$. More specifically, for each temporal frequency $f$, the respective Fourier coefficients $\hat\eta_{ki}(f)$ of region $k$ and electrode $i$ has spatial correlation: $\Cov[\hat\eta_{ki}(f), \hat\eta_{kj}(f)] = f^{-\alpha} \exp(-\frac{\mathrm{dist}^2(i,j)}{2 \sigma_\mathrm{spatial}^2})$, where $\mathrm{dist}(i,j)$ was the distance between electrodes $i$ and $j$ on one array, $\alpha = 1.4$, and
$\sigma_\text{spatial} = 0.8$.
%
Next, we let the factor loadings $\beta_{kj}$ have components
$\beta_{kji} = \gamma \cdot \exp(-\frac{\mathrm{dist}^2(i,p_{kj})}{2 \sigma_\mathrm{spatial}^2})$, $i = 1, 
\ldots d_k$, where $p_{kj}$ 
was a randomly chosen location on an array and
$\gamma$ took values in $\{0.055, 0.063, 0.071, 0.077, 0.084, 0.089\}$. This set spanned $\{0.5, 0.66, 0.84, 1.0, 1.16, 1.34\}$ times the observed signal-to-noise ratio (SNR) in the experimental data in \cref{sec:experimental_data}.
More specifically, the observed data exhibited a signal to noise ratio (defined using signal power) of approximately 0.75.
Finally, the three latent oscillations $L_{0,j}$ were set at $18$ Hz and designed to induce three epochs of lead-lag relationships between $L_1$ and $L_2$ around experimental times $80$, $200$ and $400$ ms; $L_1$ lead $L_2$ by $30$ ms during the first epoch, and
$L_1$ lagged $L_2$ by $30$ ms during the other two,
as depicted in~\Cref{fig:result_SSM}(a).

One simulated dataset consisted of $N=1000$ trials of $500$ ms long $d_1=25$ and $d_2=25$-dimensional time series $\{L_1^{(t)} \in \reals^{d_1}, L_2^{(t)} \in \reals^{d_2}\}$ generated from \cref{eq:nonstationary_lagged_mixture_model} at sampling frequency $1000$ Hz.
We filtered the simulated LFP recordings using the  complex 
Morlet wavelet at frequency $f_0 = 18$ Hz and bandwidth $50$ ms, 
the same we applied to the data in \cref{sec:experimental_data}, 
and collected the beta oscillation amplitude envelopes as the absolute values of the filtered signals. The complex Morlet wavelet is 
a complex sinusoidal  with a Gaussian envelope \citep{gabor1946theory}, where the bandwidth refers to the Gaussian standard deviation.
(The filtered amplitude at a given time $t$ is essentially the amplitude that would be obtained using a harmonic regression with cosine and sine damped by the Gaussian kernel centered at $t$; the wavelet formulation is computationally different and more efficient.)
After downsampling the power envelopes to $100$ Hz, we applied LaDynS to the resulting data $X_1$ and $X_2$ with $T = 50$ time points. 



\subsubsection{LaDynS estimation details} \label{sec:simulation_estimation}

The simulated amplitude time series were very smooth, similar to the experimental data. 
We thus added the regularizer $\lambda_\mathrm{diag}$ to the diagonal entries of the estimated correlation matrix $\hat{\Sigma}$ so it could be inverted.
Our calibration strategy for $\lambda_\mathrm{diag}$ was as follows. 
Let $X_{k,i}^{(t)}$ be an observed time series and $S_{k,i} \in \reals^{T \times T}$ be its auto-correlation matrix, $k \in \{1,2\}$, $i \in \{1, \ldots, d_k\}$.
Band-pass filtering the LFP data induced auto-correlations in $X_{k,i}^{(t)}$, which we should observe in $S_{k,i}^{-1}$, unless $S_{k,i}^{-1}$ is degenerate. 
We thus took $\lambda_\text{diag}$ to be such that $(S_{k,i} + \lambda_\text{diag} I_T)^{-1}$ displayed the expected auto-correlation. Practically, we chose $\lambda_\text{diag}$ automatically to minimize the $\ell_2$ distance between the off-diagonal entries of $(S_{k,i} + \lambda_\text{diag} I_T)^{-1}$ and the band-pass filter kernel-induced auto-correlation, after a scale adjustment, summed over $k$ and $i$.
Penalizing the diagonal introduced inevitable bias to the sparsified and desparsified LaDynS' precision estimates, $\hat{\Omega}$ and $\tilde{\Omega}$. 
The other hyperparameters were set to  
$d_\text{auto} = d_\text{cross} = 10$, $\lambda_\text{auto} = 0$ and $\lambda_\text{cross}$, the penalty on the cross-correlation elements, was determined as per \cref{sec:tuning_parameter}.

\subsubsection{Results} \label{sec:simulation_result}

\begin{figure}
    \centering
    \raisebox{4.2cm}{\bf (a)}
    \includegraphics[scale=0.6]{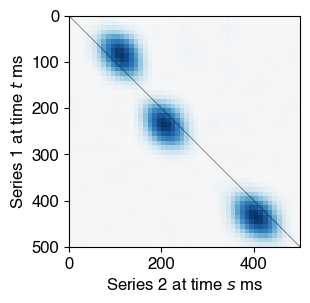}
    \quad
    \raisebox{4.2cm}{\bf (b)}
    \includegraphics[scale=0.6]{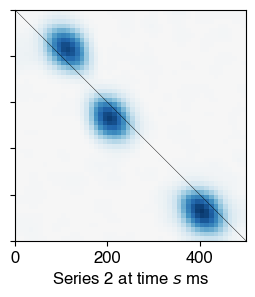}
    \quad
    \raisebox{4.2cm}{\bf (c)}
    \includegraphics[scale=0.6]{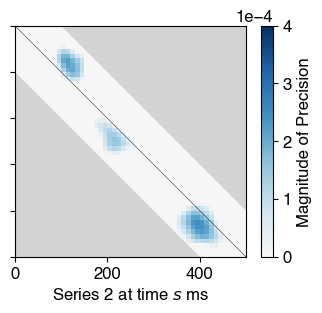}
    
    \centering
    \raisebox{4.2cm}{\bf (d)}
    \includegraphics[scale=0.6]{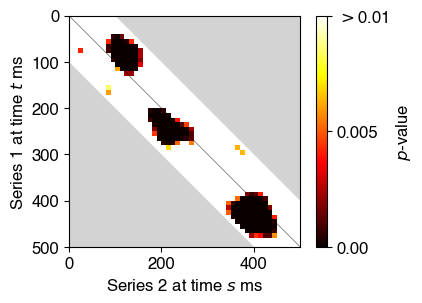}
    \quad
    \raisebox{4.2cm}{\bf (e)}
    \includegraphics[scale=0.6]{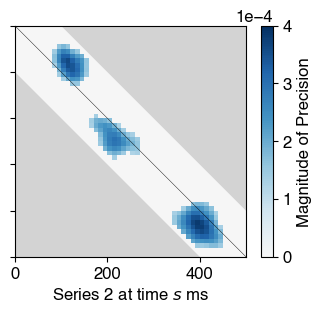}

    \caption{\sl {\bf Output and inference of LaDynS applied to simulated datasets from the shared oscillatory driver model.}
    {\bf (a)} True cross-precision matrix $\Omega_{12}$ for the connectivity scenario described in \cref{sec:simulation_from_SSM} with $\gamma=0.077$.
    {\bf (b)} Average over $60$ simulation datasets of LaDynS de-sparsified precision estimates $\tilde\Omega_{12}$. There is a good match to the true $\Omega_{12}$ in (a).
    {\bf (c)} Cross-precision estimate $\hat\Omega_{12}$ for one simulated dataset. It matches (a) up to random error.
    {\bf (d)} Permutation bootstrap p-values for the de-sparsified estimate $\tilde \Omega_{12}$.
    {\bf (e)} Discovered non-zero cross-precision estimates by the BH procedure at nominal FDR 5\%. The cluster-wise p-values of the three discovered clusters by the excursion test were all smaller than $0.5\%$. Panels (a), (b) and (c) share the same color bar in (c).}
    \label{fig:result_SSM}
\end{figure}

The simulated data true cross-precision matrix $\Omega_{12}$ is unknown
so we estimated it by simulation. 
Given a simulated dataset $(X_{1,[n]}^{(t)}, X_{2,[n]}^{(t)})$, for each trial $n=1,\dots,N$ and time $t = 1,\dots,T$, we used \cref{eq:canonical_variable} to recover the true latent factors $(Z_{1,[n]}^{(t)}, Z_{2,[n]}^{(t)})$ from 
$(X_{1,[n]}^{(t)}, X_{2,[n]}^{(t)})$ and known factor loadings $(\beta_1^{(t)}, \beta_2^{(t)})$.
We then obtained the empirical covariance matrix of the true latent factors: $\Sigma^o := \frac{1}{N} \sum Z_{[n]} Z_{[n]}^\top$, where $Z_{[n]} = (Z_1^{(1)}, \dots, Z_2^{(T)})$,
and calculated the regularized precision matrix $\Omega^o := (\Sigma^o + \lambda_\mathrm{diag} I)^{-1}$. We estimated the true precision matrix 
with the average of $200$ repeats of $\Omega^o$.
\Cref{fig:result_SSM}(a) shows the cross-regional component $\Omega^o_{12}$ of this estimate. 
We see lead-lag relationships between simulated $X_1$ and $X_2$ around $80$, $200$, and $400$ ms, as specified in \cref{sec:simulation_from_SSM}. 

\Cref{fig:result_SSM}(c) displays the LaDynS cross precision estimate $\hat\Omega_{12}$ fitted to one dataset simulated under the connectivity scenario depicted in \Cref{fig:result_SSM}(a), with connection strength $\gamma=0.077$.
\Cref{fig:result_SSM}(d) shows the permutation bootstrap p-values in \cref{eq:p_prec} (with permutation bootstrap simulation size $B = 200$) for the entries of the desparsified cross-precision estimate $\tilde\Omega_{12}$.
Small p-values concentrate near the locations of true non-zero cross-precision entries and are otherwise scattered randomly.
We then identified the significant connections by first applying the BH procedure with target FDR $5\%$  
and next by applying the excursion test at significance level 5\% to all discovered clusters (\cref{sec:inference}). 
The significant clusters are plotted in \Cref{fig:result_SSM}(e).
They match approximately the true clusters in \Cref{fig:result_SSM}(a), although they exhibit random variability, 
as we should expect. 
To average this random variability out, we estimated $\tilde{\Omega}_{12}$ for each of 60 simulated datasets,
and plotted their average in \Cref{fig:result_SSM}(b).
The average LaDynS estimate is a close match to the true cross-precision matrix in \Cref{fig:result_SSM}(a).

\begin{figure}[t!]
  \includegraphics[scale=0.6]{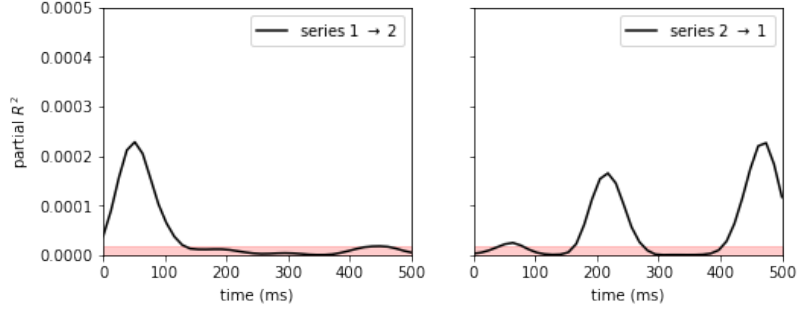}
    
  \caption{\sl
  {\bf Estimated partial $R^2$ from locally stationary state-space model} for {(left)} time series 1 $\rightarrow$ 2 and {(right)} 2 $\rightarrow$ 1. The pink shaded areas are the point-wise 95th percentiles of null partial $R^2$ under independence between the two time series.}
  \label{fig:pR2_results_SSM}
\end{figure}

\Cref{fig:pR2_results_SSM} displays the estimated partial $R^2$ 
obtained from the locally stationary state-space model described in \cref{sec:local_granger_causality}. 
The pink shaded areas are pointwise 95th percentiles of the null partial $R^2$ distribution 
under the assumption of independence between the two time series. 
The estimated effects closely reflect the true lead-lag relationships specified 
in \cref{sec:simulation_from_SSM}. Specifically, the left panel shows that time series~$1$ exerts strong Granger 
causal influence on time series~$2$ during the first connectivity epoch around $80$ ms, and the right panel 
shows that time series~$2$ Granger causes time series~$1$ during the later connectivity epochs around $200$ and $400$ ms.

\begin{figure}[t!]
    \centering
    \raisebox{4.2cm}{\bf (a)}
    \includegraphics[scale=0.6]{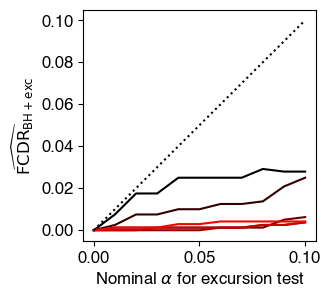}
    \quad
    \raisebox{4.2cm}{\bf (b)}
    \includegraphics[scale=0.6]{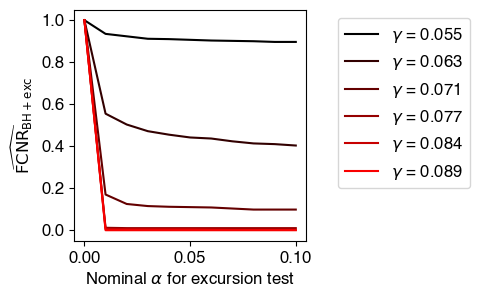}

    \caption{\sl {\bf Performance of cluster-wise inference after excursion test.}
    {\bf (a)} False cluster discovery rates and
    {\bf (b)} false cluster non-discovery rate of BH at target FDR $5 \%$ followed by excursion test at significance level $\alpha \in [0, 0.10]$ to identify non-zero partial correlations, under the connectivity scenario in \Cref{fig:result_SSM}(a), for the simulated range of connectivity intensities. 
    }
    \label{fig:fcdr_fcnr_exc_SSM}
\end{figure}

{\Cref{fig:result_SSM,fig:pR2_results_SSM} used connection strength $\gamma = 0.077$. We
repeated the simulation
for a range of $\gamma$ such that the SNR in the simulated data beta band ranged 
from zero to twice the estimated SNR of
the experimental data in \cref{sec:experimental_data}.
Instead of showing graphs we summarized the performance
of LaDynS using
false {\it cluster} discovery and non-discovery rates \citep{perone2004false} (FCDR, FCNR) in
the estimated precision matrix.
We defined a cluster of estimated nonzero precision elements to be falsely discovered if it contained no true
effect,
and a true cluster was deemed falsely non-discovered if no estimated cluster overlapped with it.  
We estimated these rates with the corresponding proportions in $60$ simulated datasets.
They are plotted in \Cref{fig:fcdr_fcnr_exc_SSM} against nominal
significance levels of the excursion test between 0 and 10\%, where the target FDR for the BH procedure was fixed at 5\%. The FCDR remained below the nominal level across all tested significance values and connectivity intensities
(all or almost all discovered clusters were true clusters) and for
large enough $\gamma$, the FCNR was zero (all clusters were 
discovered).}

{Perhaps reporting FDR and FNR would have been more informative, 
but we reported FCDR and FCNR because they were less sensitive to $\Omega_{12}$ 
being approximated rather than known in this particular example.
See \cref{app:simulation_from_model} for an example where 
$\Omega_{12}$ is known so FDR and FNR could be calculated.}



%

\noindent{\it Remark. \,\, } 
A defining feature of LaDynS is that it is capable of estimating amplitude 
lead-lag relationships between two high-dimensional oscillatory time-series 
even if there is no other source of correlations, such as coherence or phase locking,
between them.
We demonstrated this in \cref{app:simulation_no_coh} using simulated data from the model in this section, modified so it would induce amplitude lead-lag relationships without inducing phase correlations between the simulated pair of time-series. 

\begin{figure}[t!]
  \centering
    \includegraphics[scale=0.6]{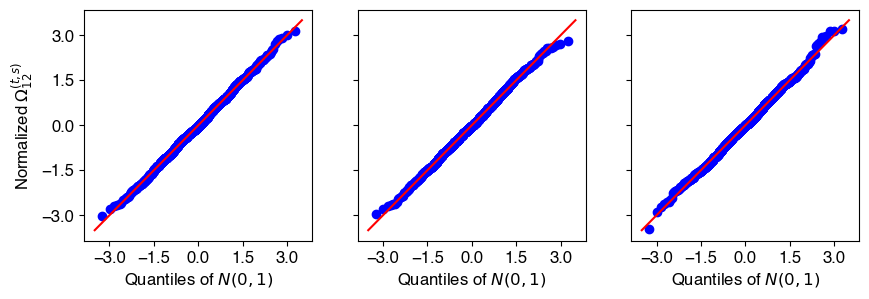}
  \caption{
  \sl {\bf Null distributions of three representative entries of $\tilde{\Omega}_{12}^{(t,s)} / \sqrt{\hat{\Var}[\tilde{\Omega}_{12}^{(t,s)}]}$}, obtained from $R = 60$ simulated datasets (\cref{sec:simulation_from_SSM}) and compared to the standard Gaussian distribution via QQ-plots. 
  The good agreement suggests that the Normal assumption in~\cref{eq:CLT_on_dspr_LaDynS} is appropriate. 
  }
  \label{fig:dspr_QQ_SSM}
\end{figure}

\begin{figure}
    \centering
        \centering
        \includegraphics[scale=0.6]{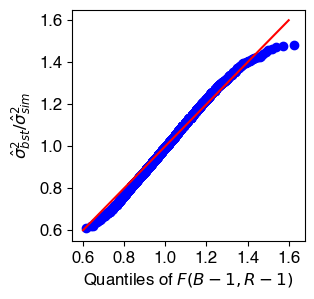} \\
    \caption{{\bf Standard deviations of desparsified precision elements.} \sl 
    F-statistics of ratios between bootstrap and empirical variances for null 
    entries of $\Omega_{12}$, showing good agreement.
    }
    \label{fig:sd_sim_vs_perm_SSM}
\end{figure}

\noindent{\it Checking the assumptions in \cref{eq:CLT_on_dspr_LaDynS,eq:p_prec}. \,\, } 
We assumed that, under the null hypothesis $H_0^{(t,s)}:\Omega_{12}^{(t,s)}=0$,
the desparsified precision entries $\tilde{\Omega}_{12}^{(t,s)}$ 
were normally distributed and
$\Var[\tilde{\Omega}_{12}^{(t,s)}]$
was well approximated by the 
permutation bootstrap variance $\hat \Var[\tilde{\Omega}_{12}^{(t,s)}]$ .
To check the former, we compared 
the empirical distribution of $R=60$ repeat estimates $\tilde{\Omega}_{12}^{(t,s)} / \sqrt{\hat{\Var}[\tilde{\Omega}_{12}^{(t,s)}]}$ to the standard normal distribution via QQ-plots.
\Cref{fig:dspr_QQ_SSM} shows QQ-plots for three randomly chosen time pairs $(t,s)$ 
that satisfy $\Omega_{12}^{(t,s)} = 0$.
The good agreement between empirical and theoretical quantiles suggests that 
the normal assumption is appropriate. 
%
Next, for each $(t,s)$ such that $H_0^{(t,s)}:\Omega_{12}^{(t,s)}=0$,
we checked the validity of the permutation bootstrap variance estimate $\hat\Var[\tilde\Omega_{12}^{(t,s)}]$
by comparing it to the empirical variance of the $R=60$ estimates $\tilde{\Omega}_{12}^{(t,s)}$.
\Cref{fig:sd_sim_vs_perm_SSM} shows the Q-Q plot of their ratios against the quantiles of 
the $F(B-1, R-1)$ distribution, which suggests that the two estimates are equal up to random
error.


\begin{figure}
    \centering
    \raisebox{3.5cm}{\bf (a)}
    \includegraphics[scale=0.5]{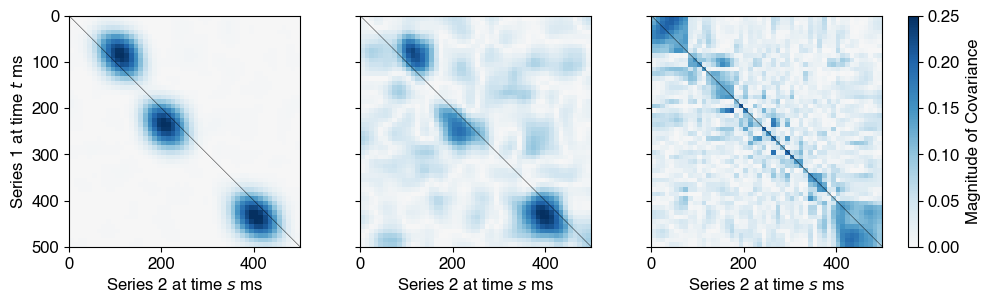}
    
    \raisebox{3.5cm}{\bf (b)}
    \includegraphics[scale=0.5]{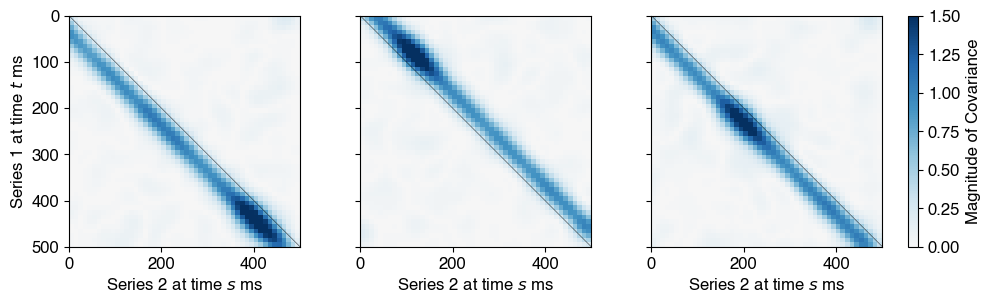}

    \raisebox{3.5cm}{\bf (c)}
    \includegraphics[scale=0.5]{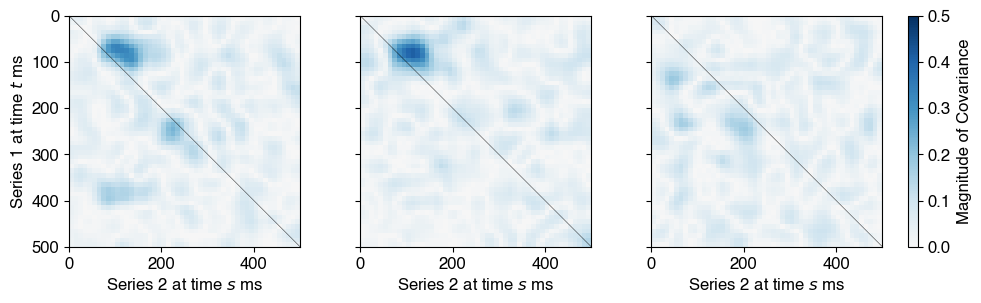}

    \raisebox{3.5cm}{\bf (d)}
    \includegraphics[scale=0.5]{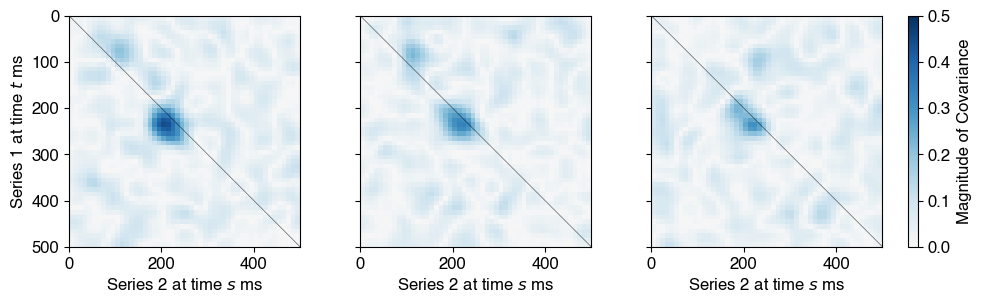}

    \raisebox{3.5cm}{\bf (e)}
    \includegraphics[scale=0.5]{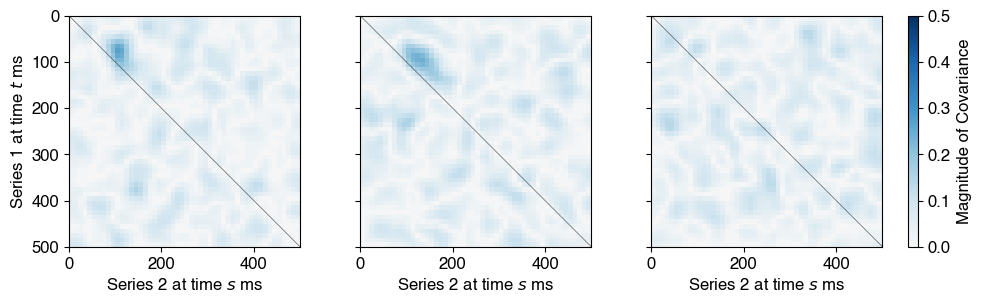}

    \caption{\sl 
    \textbf{Comparison of LaDynS and competing methods applied to one simulated dataset from the shared oscillatory driver model under the three-factor scenario.}
    \textbf{(a)} (left) The true cross-covariance matrix. 
    (center and right) Estimated cross-covariance matrices from LaDynS and DKCCA, respectively. 
    \textbf{(b)} The cross-covariance estimates obtained by DLAG, which pairs the 3 factors in region~1 with the 3 factors in region~2, yielding three cross-covariance plots (one for each factor pair). 
    \textbf{(c-e)} The cross-covariance estimates obtained by GPFA, which considers all possible between-region factor pairs among the 3 factors in region~1 and the 3 factors in region~2, yielding nine cross-covariance plots.}
    \label{fig:comparison_1}
\end{figure}

\begin{figure}
    \centering
    \raisebox{3.5cm}{\bf (a)}
    \includegraphics[scale=0.5]{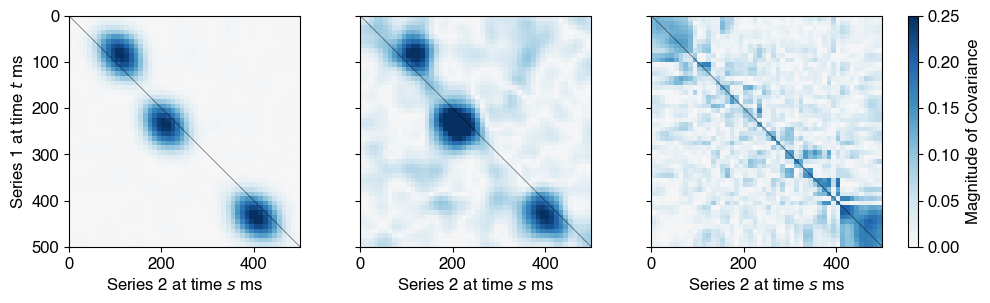}
    
    \raisebox{3.5cm}{\bf (b)}
    \includegraphics[scale=0.5]{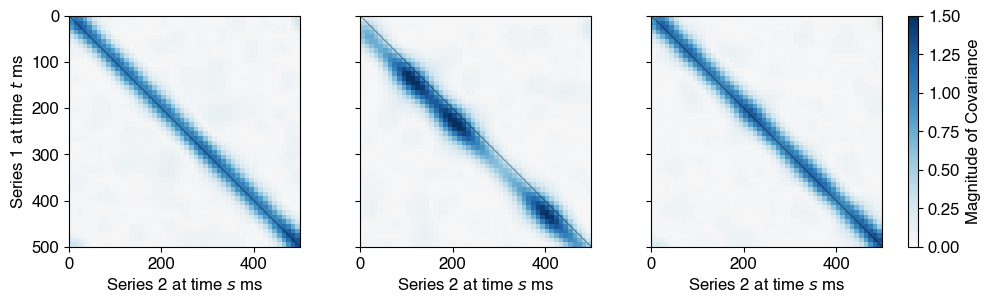}

    \raisebox{3.5cm}{\bf (c)}
    \includegraphics[scale=0.5]{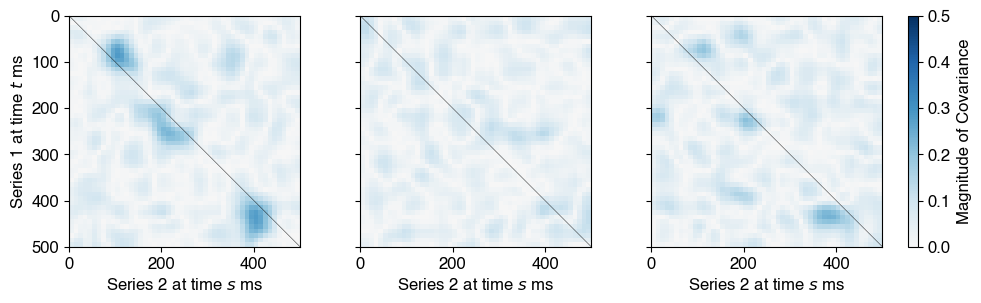}

    \raisebox{3.5cm}{\bf (d)}
    \includegraphics[scale=0.5]{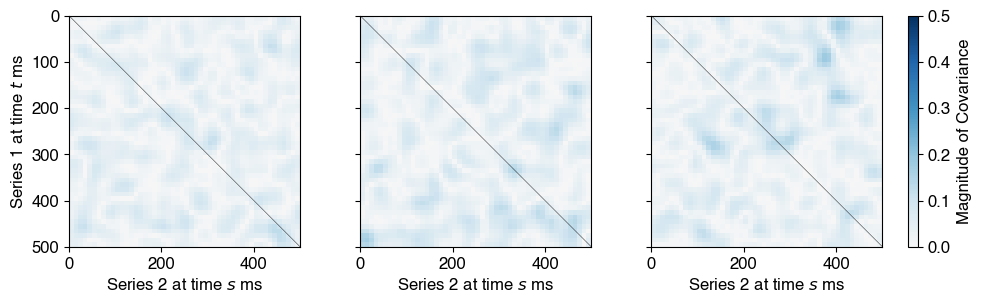}

    \raisebox{3.5cm}{\bf (e)}
    \includegraphics[scale=0.5]{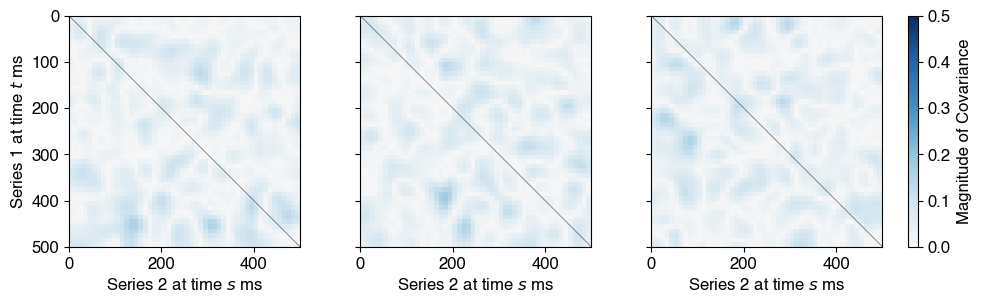}

    \caption{\sl
    \textbf{Comparison of LaDynS and competing methods applied to one simulated dataset from the shared oscillatory driver model under the single-factor scenario.}
    \textbf{(a)} (left) The true cross-covariance matrix. 
    (center and right) Estimated cross-covariance matrices from LaDynS and DKCCA, respectively. 
    \textbf{(b)} The cross-covariance estimates obtained by DLAG, which pairs the 3 factors in region~1 with the 3 factors in region~2, yielding three cross-covariance plots (one for each factor pair). 
    \textbf{(c-e)} The cross-covariance estimates obtained by GPFA, which considers all possible between-region factor pairs among the 3 factors in region~1 and the 3 factors in region~2, yielding nine cross-covariance plots.}
    \label{fig:comparison_2}
\end{figure}

{
\subsection{Comparison to existing methods}
\label{sec:comparison}

In this section we compare the performance of LaDynS on simulated data generated from the shared oscillatory driver model (see \cref{sec:simulation_from_SSM}) with three existing methods: Dynamic Kernel Canonical Correlation Analysis (DKCCA, \citealp{rodu2018detecting}), 
Delayed Latents Across Groups (DLAG, \citealp{gokcen2021disentangling}), and Gaussian Process Factor Analysis (GPFA, \citealp{yu2009gaussian}). The simulated data are generated under two distinct connectivity scenarios:
\begin{itemize}
    \item \emph{Three-factor scenario}: Three pairs of latent factors drove the connections, each active in a distinct time epoch, as depicted by the true cross-covariance matrix in \Cref{fig:comparison_1}(a), left panel. Each pair had a unidirectional influence with a fixed lead-lag.

    \item \emph{Single-factor scenario}: The epochs and lead-lags of the connections were the same as in the three-factor scenario (\Cref{fig:comparison_2}(a), left panel), but this time a single pair of latent factors drove all the three epochs of connections. As a result, the connection between the pair of latent factors was bidirectional, and the lead-lag changed over time.
\end{itemize}

LaDynS models each latent factor time series \(Z_k^{(t)}\) to have time-varying factor loadings \(\beta_k^{(t)}\), accommodating changes in the activated latent factors across time \(t\). Similarly, DKCCA allows each canonical component time series to have time-varying canonical weights. As a result, both LaDynS and DKCCA can capture multiple factor-driven connections using only a single pair of latent factors or canonical components, provided these underlying factors are active in distinct time intervals (e.g., the three-factor scenario). By contrast, DLAG and GPFA use fixed factor loadings over time, and thus require three latent factors per set of observed time series to capture the same three-factor scenario.
Accordingly, we estimated the interaction between two sets of simulated time series, \(X_1^{(t)}\) and \(X_2^{(t)}\), under the following configurations:
\begin{itemize}
    \item \textbf{LaDynS:} We followed the estimation details in \cref{sec:simulation_estimation}. Although LaDynS represents the between-sets interaction via the cross-precision matrix \(\Omega_{12}\), we use the cross-covariance matrix \(\Sigma_{12}\) here to facilitate comparison with the other methods, which estimate cross-covariance matrices only.

    \item \textbf{DKCCA:} We estimated the cross-covariance between a single pair of canonical components defined by a sliding time window of width 9 (\(g = 4\) in the original notation).

    \item \textbf{DLAG:} We used 3 between-group and 2 within-group factors for each set of time series. In DLAG, each between-group factor in one set is exclusively paired with a counterpart in the other set, yielding 3 cross-covariance matrix estimates.

    \item \textbf{GPFA:} Because GPFA inherently models a single set of time series, it does not directly provide a cross-connection component. We therefore used a two-step approach: first, we fitted GPFA with 3 latent factors to each of \(X_1^{(t)}\) and \(X_2^{(t)}\); then, we computed cross-covariance matrices for all \(3 \times 3\) pairs of latent factors, producing 9 cross-covariance estimates.
\end{itemize}
For all other tuning hyperparameters of DKCCA, DLAG, and GPFA, we used the default settings provided in the example scripts of the respective code packages. For details, we refer the reader to the example code in our \texttt{ladyns} package.}




\Cref{fig:comparison_1} presents the simulation results for the three-factor scenario. Both LaDynS and DLAG successfully identified all three interaction epochs (\Cref{fig:comparison_1}(a), center panel, and (b)), whereas GPFA only captured the first two (\Cref{fig:comparison_1}(c-e)), and DKCCA failed to detect the interaction entirely (\Cref{fig:comparison_1}(a), right panel). This outcome for GPFA is unsurprising, given that GPFA was designed primarily to capture dominant within-set dynamics rather than cross-set interactions. Consequently, if cross-regional interactions are relatively weak compared to the within-region variability (as in this three-factor scenario), GPFA may fail to detect all active connections.

Under the single-factor scenario, where the interaction between a single pair of latent factors is sufficiently strong, GPFA was able to recover all three interaction epochs (\Cref{fig:comparison_2}(c-e)), matching the performance of LaDynS (\Cref{fig:comparison_2}(a), center panel). By contrast, DLAG struggled to capture the bidirectional nature of the interaction (\Cref{fig:comparison_2}(b)), consistent with its assumption that each pair of between-group latent factors has a fixed lead-lag relationship. 
We additionally ran DLAG with a single between-group latent factor to match the true latent dimensionality in this scenario; however, its qualitative behavior remained unchanged, and DLAG continued to struggle to capture the bidirectional nature of the interaction.
DKCCA again did not detect any meaningful interaction in this setting.

Overall, LaDynS was the only method that consistently identified all three interaction epochs 
in both connectivity scenarios. DKCCA showed weak performance in both scenarios, while DLAG 
and GPFA had diminished power in at least one of the scenarios. The performances of DLAG 
and GPFA would further degrade if we did not specify the correct number of latent variables.

\subsection{Experimental Data Analysis} \label{sec:experimental_data}


\begin{figure}[t!]
    \centering
    \raisebox{4.2cm}{\bf (a)}
    \includegraphics[scale=0.6,trim={.2cm 0 0 0},clip]{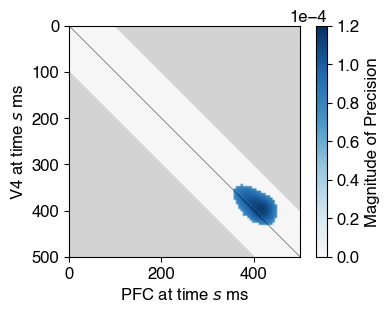}
    \raisebox{4.2cm}{\bf (b)}
    \includegraphics[scale=0.6,trim={.2cm 0 0 0},clip]{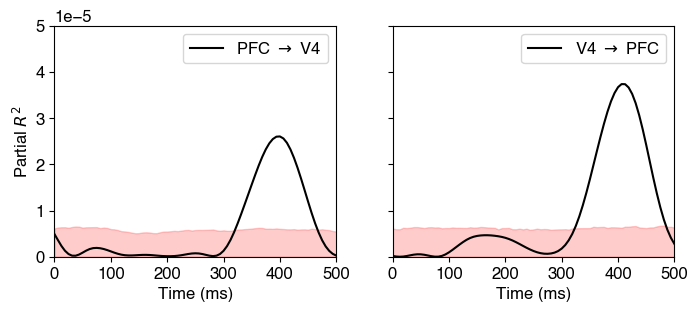}

    \caption{\sl {\bf Experimental data analysis.
    (a) LaDynS inference.} Discovered region of cross-precision using BH at nominal FDR 5\% and the excursion test ($p < 0.0005$). The light gray area shows the region of time considered (one area leading the other by at most $100$ ms).
    The blue blob suggests that activity in V4 preceded that in PFC around $400$ ms in the delay period. 
    {\bf (b) Estimated partial $R^2$ from locally stationary state-space model} for {(left)} PFC $\rightarrow$ V4 and {(right)} V4 $\rightarrow$ PFC. The pink shaded areas 
    indicate the range of values that fall below the  95th percentile under the null hypothesis of independence between V4 and PFC.}
    \label{fig:result_Smith}
\end{figure}

We applied LaDynS to local field potentials (LFPs), collected in the experiment described in \citet{khanna2019dynamic}, from two Utah arrays implanted in a Macaque monkey's prefrontal cortex (PFC) and V4 during a memory-guided saccade task. Each trial of the task started with a monkey fixating its eyes {on a dot} at the center of the screen. A visual cue was given for $50$ ms to indicate a target location, which was randomly chosen from $40$ {candidate locations (8 directions and 5 amplitudes)} that tiled the display screen. The cue was turned off and the monkey had to remember the target location while maintaining eye fixation for a delay period of $500$ ms. After the delay period, the monkey {made a saccade to the} remembered position, and {a liquid reward was provided} on successful trials. As in \citet{khanna2019dynamic}, 
we analyzed the time series during the delay period,
based on $3000$ successful trials.  Time $t=0$ corresponds to 
the start of that period. The data are available in \citet{snyder2022utah}.
%
Because beta oscillations are often associated with communication across brain areas \citep{klein2020torus,miller2018wm2},
we filtered LFP recordings 
at a beta oscillation frequency $18$ Hz and obtained the beta oscillation power envelopes 
as described in \cref{sec:simulation_estimation}. We chose $18$ Hz because it was the frequency having the largest power within the range 12-40 Hz (see \Cref{fig:spectrogram_Smith}).
After downsampling the power envelopes to $200$ Hz, we applied LaDynS with the regularizer $\lambda_\text{diag}$ on the diagonal of $\Sigma$, as in \cref{sec:simulation_estimation}, 
because the filtered data were very smooth.


\Cref{fig:result_Smith}(a) shows the only epoch of significant 
contiguous region of the precision matrix identified by our method 
(FDR at 5\%, $p < .0005$ by the excursion test in \cref{sec:inference}).
This result provides strong evidence that the $18$ Hz beta amplitudes in PFC and V4 were correlated (after conditioning on beta amplitudes at all other times and lags) around 400 milliseconds after the start of the delay period.

\begin{figure}[t!]

  \includegraphics[scale=0.6]{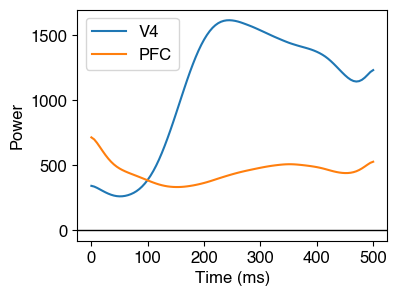}
  \caption{\sl {\bf Estimated beta powers of electrophysiological activity driven by latent factors in V4 and PFC as functions of time}. 
  The summed beta power in the data $X_k^{(t)}$ attributable to the latent time series at time $t$ was estimated by the $\ell_2$ norm of the factor loading vector $\beta_k^{(t)}$; see \cref{eq:pCCA_obs_over_time}.
  }
  \label{fig:fnorm_Smith}
\end{figure}

To better understand this  relationship, we used the estimated latent time series to compute partial $R^2$ values under an assumption of locally stationarity, applying the model in \cref{eq:AR_latent_model}, \cref{sec:local_granger_causality}.
The frequency $18$ Hz corresponds to a period of $55$ ms and it is hard to detect non-stationarity 
at time scales finer than a few periods. We 
therefore used a moving window of $100$ ms (i.e. larger than the period) to calculate partial $R^2$, allowing connection delays between
V4 and PFC in the $\tau_1=15$ to $\tau_2=30$ ms range.
The partial $R^2$ from PFC to V4 and from V4 to PFC are shown in \Cref{fig:result_Smith}(b).
There are large excursions of $R^2$ above the null values in both plots, suggesting that, at around 400 ms post stimulus, the two areas are involved in a bidirectional network: the power of the activity in each area predicts the power of the oscillation in the other, following a short delay (after conditioning on the power at all other times and lags).
To see that this is not due solely to the activity passing through the visual stream, in \Cref{fig:fnorm_Smith}, for each of the two latent time series, we plotted the estimated total beta power as a function of time. 
The power in V4 increases dramatically much earlier than 400 ms, starting around 100 ms and reaching a peak just after 200 ms; it then remains substantial during the remainder of the delay period. 
The results in \Cref{fig:result_Smith} can not be explained by those in 
 \Cref{fig:fnorm_Smith} alone.

\begin{figure}[h!]
  \centering \setlength{\labelsep}{0mm}
  \includegraphics[scale=0.6]{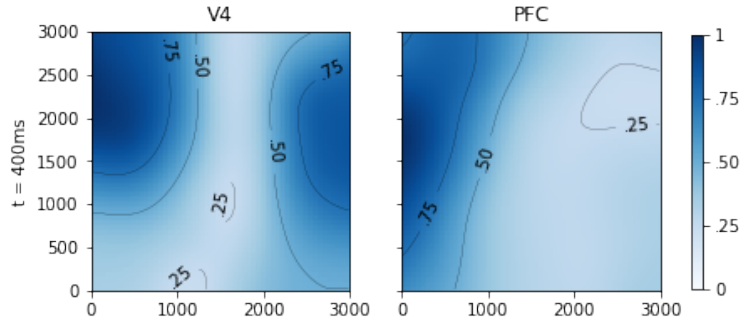}
  
  \caption{\sl {\bf Factor loadings of V4 and PFC}, spatially smoothed, normalized, and color coded over the electrode coordinates ($\mu m$) on {\bf (left)} V4 and {\bf (right)} PFC at late delay period ($400$ ms).
  Contours at .25, .5 and .75 
  of the maximal power were added.}
  \label{fig:beta_results}
\end{figure}

It is also possible to get spatial information from the normalized factor loadings across the electrode arrays, which are displayed in \Cref{fig:beta_results} for experimental time $400$ ms. The loadings were rescaled by their maximal value across the array. 
(An animation over the complete timeline is available at \url{github.com/HeejongBong/ladyns}.) It is apparent that a relatively small proportion of the electrodes, residing in limited portions of the recording areas, contribute most of the beta oscillatory power identified by the bivariate time series.

%% file: sections/4_discussions.tex
\section{Discussion} \label{sec:4_conclusion}

{To describe cross-population interactions of oscillatory amplitudes in high-dimensional neural field potentials we developed a novel and intuitive procedure: in LaDynS, cross-population interactions are described using the cross-correlation of latent drivers. To arrive at a statistically rigorous approach,} we provided a time-series extension of probabilistic CCA together with a novel sparse estimation methodology. According to our Equation (\ref{eq:pCCA_obs_over_time}), each of the two multivariate time series is driven by a single latent time series, with the cross-dependence of these two latent time series representing cross-region interaction. According to Equation (\ref{eq:pCCA_latent_over_time}), the latent bivariate time series is a discrete Gaussian process but its correlation matrix is unrestricted, allowing for {time-varying dynamics, i.e., statistical non-stationarity}. The repeated trial structure enabled us to estimate the resulting high-dimensional covariance matrix by applying sparse estimation and inference methods.
We found, and displayed in \Cref{fig:result_Smith}(b), an interesting interaction between PFC and V4, involving beta power, that appeared during late delay period, 
with the interaction being bidirectional. 
The results were based on partial $R^2$ values, computed from the estimated covariance matrices, corresponding to lagged regressions of one latent time series on the other. 
The analysis in \Cref{fig:result_Smith}(b) is in the spirit of Granger causality, but differs from it by allowing for non-stationarity, so that we could obtain the time-varying results. 

The partial $R^2$ values in \Cref{fig:result_Smith}(b) are highly statistically significant, but they are very small. We note that our diagonal regularization of the estimated latent precision matrix artificially reduced these values, so that the scale is no longer interpretable in familiar terms. The cross-correlation estimates are several orders of magnitude larger (see \Cref{fig:Sigma_Smith}). However, they remain smaller than .1, as are the raw correlations at 400 ms and 425 ms, respectively, across individual electrodes in V4 and PFC (not shown).
This reflects the highly inhomogeneous nature of the LFP voltage recordings together with the dominance of low frequencies (unfiltered LFP spectra typically follow $1/f^\alpha$ trends, where $f$ is frequency and $\alpha$ is roughly 1.5 to 2; {see \citealp{kass2023identification}}). The large number of repeated trials allowed us to find and document the results.

In addition to making the analysis possible, the repeated trial structure suggests substantive interpretation based on trial-to-trial variability. Although investigators take pains to make the experimental setting nearly the same on each trial, the inevitable small fluctuations in the way the subject interacts with the environment, together with changes in the subject's underlying state (involving alterations in motivational drive, for example), lead to observable fluctuations in behavior and in the recorded neural activity. Although the network sources of trial-to-trial variability in the PFC and V4 data are unknown, they produce the kind of correlated activity revealed in \Cref{fig:result_Smith}(b). 
To interpret it, we acknowledge there could be some potentially confounding trial-to-trial variation --- such as changing motivational state, stimulus timing, bandpass-filter distortions and volume conduction --- that drive beta power
in V4 and PFC, having just the right differential time lags to produce the correlated activity picked up by the partial $R^2$ values, as cautioned by \citet{lintas2021operant}.
Could such task-irrelevant pulses of activity change across time, within repetitions of the task, in such a way as to produce, the peaks in \Cref{fig:result_Smith}(b)?
It is possible, but it would be surprising, especially when we consider contemporary ideas about beta oscillations during working memory tasks \citep{miller2018wm2} along with the well-identified distinction between early and late visual processing that, presumably, corresponds to a distinction between feedforward and recurrent (bidirectional) flow of information  \citep{super2001two,buschman2007topdown, del2007brain,yang2019exploring,chen2022population,mejias2016feedforward}. The alternative we mentioned, that PFC and V4 are involved, together, in goal-directed visual processing and memory, with the two areas acting bidirectionally around 400 ms post stimulus, seems  likely.

There are many ways to extend the ideas developed here.
While we applied LaDynS to LFP recordings, the framework is applicable more broadly to other slowly varying multidimensional time series. In particular, the proposed methodology can be extended to other neural modalities such as EEG, MEG, and fMRI. Although data from these modalities often violate global stationarity assumptions underlying classical approaches such as cross-correlograms \citep{guan2020profiles,klonowski2009everything,chang2010time}, local stationarity remains reasonable when functional connectivity depends on cognitive/behavioral states which evolve on time scales longer than the sampling interval. While we have not yet explored such extensions empirically, LaDynS is particularly appealing in this regime, as it explicitly accommodates slowly varying connectivity and cross-correlation over time. At the same time, signals in EEG and MEG often reflect mixtures of activity from multiple brain structures and switches between metastable neural states, which may introduce additional confounding and complicate the interpretation of inferred associations.

LaDynS may perform less favorably in settings where within-region noise autocorrelations dominate the signal. The simulation results in Figure 4 of \citet{bong2020latent} illustrate the resulting loss of specificity under such adversarial conditions. In these cases, methods that explicitly model within region autocorrelated noise, such as DLAG \citep{gokcen2021disentangling}, can outperform LaDynS.
In \citet{bong2020latent}, we proposed an extension of LaDynS that addresses this issue by allowing the within region noise processes $\epsilon_k$ to follow general time series structures and by modeling the latent processes driving each brain region as multidimensional. That brief report, which focused on data filtered at a different frequency band, did not provide the methodological details or inferential procedures developed here. An important direction for future work is to generalize the present framework to the setting considered in \citet{bong2020latent}.

Another important direction for future research is a theoretical analysis of how regularization hyperparameters affect recovery of the underlying lag structure. In this work, we focused on proposing a regularization based approach to functional connectivity estimation together with a data driven hyperparameter calibration scheme and demonstrated its performance through simulation studies, while establishing theoretical guarantees for hyperparameter selection and optimality remains an interesting topic for future investigation.
In addition, developing theory that accounts for dependence across experimental trials is of particular interest, as such dependence may arise from long term memory or adaptation effects in neural recordings and could impact the validity of the proposed inference procedure.

Furthermore, for band-pass filtered data, such as those analyzed in \cref{sec:experimental_data}, phase analysis \citep{klein2020torus} could be combined with amplitude analysis based on the complex normal distribution, as in \cite{urban2023oscillating} (which developed a latent variable model at a single time point).
Multiple frequencies and nonlinear association, such as cross-bicoherence, could be considered, along the lines of \citet{tort2010measuring,gallagher2017cross,aliramezani2024delta,abe2024detection}, as well.
A different direction for additional research would be to simplify the version of LaDynS we have used here by imposing suitable spatiotemporal structure on the latent time series. Regardless of whether such approaches are fruitful, the general framework of LaDynS could be of use whenever interest focuses on {time-varying} interactions among groups of repeatedly-observed multivariate {neural} time series.

%% file: appendices/a_proofs.tex


\section{Proof of Theorems~\ref{thm:equivalence_to_CCA} and~\ref{thm:equivalence_to_genvar}}\label{app:pf_thm_equiv_to_genvar}


In the model of \cref{eq:pCCA_obs_over_time,eq:pCCA_latent_over_time}, the marginal covariance matrix $S$ of has sub-matrices $S_{kl}^{(t,s)} = \beta_k^{(t)} \Sigma_{kl}^{(t,s)} \beta_l^{(s)\top} + \Phi_k^{(t)} \delta_{kl}^{(t,s)}$, where $
\delta_{kl}^{(t,s)} =  1$ if $k = l \textand t=s$ and $0$ otherwise, for $t, s \in [T]$ and $k, l = 1, 2$.
Let $u_k^{(t)} = S_{kk}^{(t,t)-\frac{1}{2}} \beta_k^{(t)}$ and $\Psi_k^{(t)} = S_{kk}^{(t,t)-\frac{1}{2}} \Phi_k^{(t)} S_{kk}^{(t,t)-\frac{1}{2}}$.
We note that 
$u_k^{(t)\top} u_k^{(t)} = \beta_k^{(t)\top} S_{kk}^{(t,t)-1} \beta_k^{(t)} \leq 1$. Because $\beta_k^{(t)^\top} S_{kk}^{(t,t)-1} \beta_k^{(t)}$ is non-identifiable, by adjusting $\Sigma_{k,l}^{(t,s)}$ and $\Phi_k^{(t)}$ for $(k,t) \neq (l,s)$, we can assume $\beta_k^{(t)^\top} S_{kk}^{(t,t)-1} \beta_k^{(t)}=1$.
Then
\begin{eqnarray*}
    u_k^{(t)\top} \Psi_{kk}^{(t,t)} u_k^{(t)} & = & \beta_k^{(t)\top} S_{kk}^{(t,t)-1} \Phi_k^{(t)} S_{kk}^{(t,t)-1} \beta_k^{(t)} \\
    & = & \beta_k^{(t)\top} S_{kk}^{(t)-1} (S_{kk}^{(t)} - \beta_k^{(t)} \beta_k^{(t)\top}) S_{kk}^{(t)-1} \beta_k^{(t)} \\
    & = & \beta_k^{(t)\top} S_{kk}^{(t)-1} \beta_k^{(t)} - (\beta_k^{(t)\top} S_{kk}^{(t)-1} \beta_k^{(t)})^2 \\
    & = & 1 - 1^2 = 0,
\end{eqnarray*}
for $t \in [T]$ and $k \in \{1,2\}$. That is, $u_k^{(t)}$ is orthogonal to  $\Psi_{kk}^{(t,t)}$.

Denoting the block diagonal matrix consisting of $\{S_{kk}^{(t,t)}: ~ t \in [T], ~ k \in \{1,2\}\}$ by $V$, 
$R = V^{-\frac{1}{2}} S V^{-\frac{1}{2}}$ consists of sub-matrices 
\begin{eqnarray*}
    R_{kl}^{(t,s)} & = & S_{kk}^{(t,t)-\frac{1}{2}} S_{kl}^{(t,s)} S_{ll}^{(s,s)-\frac{1}{2}} = u_k^{(t)} \Sigma_{kl}^{(t,s)} u_l^{(s)\top} + \Psi_k^{(t)} \delta_{kl}^{(t,s)}.
\end{eqnarray*}
Due to the orthogonality between $u_k^{(t)}$ and $\Psi_k^{(t)}$, $\det(R) = \det(\Omega) / \prod_{k,t} \mathrm{pdet}(\Psi_k^{(t)})$, and $Q = R^{-1}$ consists of sub-matrices
\begin{eqnarray*}
    Q_{kl}^{(t,s)} = u_k^{(t)} \Omega_{kl}^{(t,s)} u_l^{(s)\top} + \Psi_k^{(t)+} \delta_{kl}^{(t,s)},
\end{eqnarray*}
where $\Omega = \Sigma^{-1}$ is the precision matrix and $\mathrm{pdet}(A)$ and $A^{+}$ are the pseudo-determinant and Moore-Penrose pseudo-inverse of a positive semi-definite matrix $A$. Notice that $\Psi_k^{(t)} = I - u_k^{(t)} u_k^{(t)\top} = \Psi_k^{(t)+}$ and hence $\mathrm{pdet}(\Psi_k^{(t)}) = 1$. In turn, the negative log-likelihood under the model (\cref{eq:pCCA_obs_over_time,eq:pCCA_latent_over_time}) 
given observed time-series $\{X_{1,[n]}, X_{2,[n]}\}_{n = 1, \dots, N}$ is
\begin{equation} \label{eq:nll}
\begin{split}
    & \mathrm{nll}(\Sigma, \{\mu_k^{(t)}, \beta_k^{(t)}, \Phi_k^{(t)}\}_{(k,t)}; \{X_{1,[n]}, X_{2,[n]}\}_{n=1,\dots,N}) \\
    & = - \log\det(\Omega) + \sum_{k,t} \log\mathrm{pdet}(\Psi_k^{(t)}) + \sum_{k,t} \log\det(S_{kk}^{(t,t)})  \\
    & \quad + \tr(\Omega \bar\Sigma) + \sum_{k,t} \tr(\Psi_k^{(t)+} S_{kk}^{(t,t)-\frac{1}{2}} \bar S_{kk}^{(t,t)} S_{kk}^{(t,t)-\frac{1}{2}}) \\
    & = - \log\det(\Omega) + \tr(\Omega \bar\Sigma)
    + \sum_{k,t} \left\{\log\det(S_{kk}^{(t,t)}) + \tr((S_{kk}^{(t,t)-1} - w_k^{(t)} w_k^{(t)\top}) \bar S_{kk}^{(t,t)}) \right\} 
\end{split}
\end{equation}
where $\bar\Sigma_{kl}^{(t,s)} = w_k^{(t)\top} \bar S_{kl}^{(t,s)} w_l^{(s)}$, $\bar S_{kl}^{(t,s)} = \bar\Exp[(X_k^{(t)} - \mu_k^{(t)})(X_l^{(s)} - \mu_l^{(s)})^\top]$ and $w_k^{(t)} = S_{kk}^{-1/2} u_k$ for $t, s \in [T]$ and $k, l \in \{1, 2\}$, and $\bar\Exp$ indicates the sample variance mean. 
Due to the first-order optimality with respect to $\mu_k^{(t)}$, $\mu_k^{(t)} = \bar\Exp[X_k^{(t)}]$ for $k \in \{1,2\}$ and $t \in [T]$.
On the other hand, due to the first-order optimality 
with respect to $S_{kk}^{(t,t)-1}$, for all $k \in \{1, 2\}$ and $t \in [T]$,
\begin{eqnarray*}
    \nabla_{S_{kk}^{(t,t)-1}} \mathrm{nll} 
    & = & S_{kk}^{(t,t)} - \bar{S}_{kk}^{(t,t)} = S_{kk}^{(t,t)} w_k^{(t)} \lambda_k^{(t)} w_k^{(t)\top} S_{kk}^{(t,t)},
\end{eqnarray*}
where $\lambda_k^{(t)} \in \reals$ is the Lagrange multiplier corresponding to $w_k^{(t)\top} S_{kk}^{(t,t)} w_k^{(t)} = 1$. Because $w_k^{(t)\top} S_{kk}^{(t,t)} w_k^{(t)} = 1$,
\begin{equation*}
\begin{aligned}
    1 - w_k^{(t)\top} \bar{S}_{kk}^{(t,t)} w_k^{(t)} 
    & = w_k^{(t)\top} S_{kk}^{(t,t)} w_k^{(t)} - w_k^{(t)\top} \bar{S}_{kk}^{(t,t)} w_k^{(t)} \\
    & = w_k^{(t)\top} S_{kk}^{(t,t)} w_k^{(t)} \lambda_k^{(t)} w_k^{(t)\top} S_{kk}^{(t,t)} w_k^{(t)} 
    = \lambda_k^{(t)}.
\end{aligned}
\end{equation*}
Therefore, the two terms $\log\det(S_{kk}^{(t,t)})$ and $\tr((S_{kk}^{(t,t)-1} - w_k^{(t)} w_k^{(t)\top}) \bar S_{kk}^{(t,t)})$ are rewritten by
\begin{equation*}
\begin{aligned}
    \log\det(S_{kk}^{(t,t)}) 
    & = - \log(1-\lambda_k^{(t)}) + \log\det(\hat{S}_{kk}^{(t,t)}), \\ 
    \tr((S_{kk}^{(t,t)-1} - w_k^{(t)} w_k^{(t)\top}) \bar S_{kk}^{(t,t)}) 
    & = \tr((S_{kk}^{(t,t)-1} - w_k^{(t)} w_k^{(t)\top})(S_{kk}^{(t,t)} - S_{kk}^{(t,t)} w_k^{(t)} \lambda_k^{(t)} w_k^{(t)\top} S_{kk}^{(t,t)})) \\
    & = d_k - 1\\
\end{aligned}
\end{equation*}
%
Plugging it to \cref{eq:nll}, the MLE reduces to minimizing
\begin{equation*}
    \mathrm{nll}(\Omega, w_k^{(t)}, \lambda_k^{(t)}; \{X_{1,[n]}, X_{2,[n]}\}_{n=1,\dots,N}) 
    = - \log\det(\Omega) - \sum_{k,t} \log(1-\lambda_k^{(t)}) + \tr(\Omega \bar\Sigma) 
\end{equation*}
such that $\diag(\Omega^{-1}) = \mathbf{1}$. Let $w_k'^{(t)} = w_k^{(t)}/\sqrt{1-\lambda_k^{(t)}}$, $\Omega' = \diag(\sqrt{1-\lambda_k^{(t)}}) ~\Omega ~\diag(\sqrt{1-\lambda_k^{(t)}})$, and $\bar\Sigma' = \bar{\Var}[(w_1'^{(1)\top} X_1^{(1)}, \dots, w_2'^{(T)\top} X_2^{(T)})]$. Note that $\diag(\bar\Sigma') = \mathbf{1}$. The objective is rewritten by
\begin{equation*}
    \mathrm{nll}(\Omega', w_k'^{(t)}; \{X_{1,[n]}, X_{2,[n]}\}_{n=1,\dots,N}) 
    = - \log\det(\Omega') + \tr(\Omega' \bar\Sigma') 
\end{equation*}
which is maximized when $\Omega' = \bar{\Sigma}^{'-1}$ given $w_k'^{(t)}$'s are fixed.
sThus, the maximum likelihood estimation is equivalent to finding $w_k'^{(t)}$ minimizing $\log\det(\bar\Sigma')$ under $w_k'^{(t)\top} \bar{S}_{kk}^{(t,t)} w_k'^{(t)} = 1$ for $k \in [K]$, which is the GENVAR procedure of \citet{kettenring1971canonical}. This proves the desired results for the MLE with $m_k^{(t)} = 1$. The other MLEs with $m_k^{(t)} < 1$ correspond to the non-identifiable parameter sets, which have $u_k^{(t)\top} u_k^{(t)} < 1$.

\cref{thm:equivalence_to_CCA} is a corollary of \cref{thm:equivalence_to_genvar}. To see that,
let $T=1$ so that $X_1 \equiv X_1^{(1)}$, $X_2 \equiv X_2^{(1)}$ and
$
    \Sigma = \begin{pmatrix} 1 & \Sigma_{12}^{(1,1)} \\ \Sigma_{12}^{(1,1)\top} & 1 \end{pmatrix} = \begin{pmatrix} 1 & \sigma_{12} \\ \sigma_{12} & 1 \end{pmatrix}
$.
The \texttt{GENVAR} procedure solves
\begin{equation*}
    \argmin_{w_1, w_2} \det \left(\overline{\Var}\left[w_1^\top X_1, w_2^\top X_2\right] \right) \equiv \argmin_{w_1, w_2} \det\left(\begin{pmatrix} 1 & \bar{\sigma}_{12} \\ \bar{\sigma}_{12} & 1 \end{pmatrix}\right), 
\end{equation*}
where $\bar{\sigma}_{12} = \frac{w_1^T \bar{\Sigma}_{12} w_2}{\sqrt{w_1^T \bar{\Sigma}_1 w_1} \sqrt{w_2^T \bar{\Sigma}_2 w_2}}$. 
This minimization problem is equivalent to the CCA problem in \cref{eq:CCA} and
$%
    \overline{\Var}\left[\hat{w}_1^T X_1, \hat{w}_2^T X_2\right] = \begin{pmatrix} 1 & \hat{\sigma}_{cc} \\ \hat{\sigma}_{cc} & 1 \end{pmatrix},
$ 
which implies $\hat{\beta}_k = \bar{\Sigma}_{kk} \hat{w}_k m_k$ and $m_1 m_2 \hat{\sigma}_{12} = \hat{\sigma}_{cc}$ for $\abs{m_k} \leq 1$, as in \cref{thm:equivalence_to_CCA}. 

%% file: appendices/b_algorithm.tex
\section{Fitting LaDynS}

\subsection{Coordinate Descent Algorithm} \label{app:LaDynS_algorithm}

\begin{algorithm}[h]
\caption{Coordinate descent algorithm to fit LaDynS}
\begin{algorithmic}[1]
    \INPUT \hfill\\
    $\{X_k: k = 1, \dots, K\}$: input data \\
    $\Lambda \in [0,\infty]^{KT \times KT}$: sparsity penalty matrix \\
    $\text{iter}_\text{max} \in \nats_{+}$: maximum iteration \\
    $\text{ths} \in \reals_{+}$: threshold for convergence
    \\\hfill\\
    \OUTPUT $\Omega$ and $w_k^{(t)}$'s which solve \cref{eq:penalized_likelihood} w.r.t. \cref{fig:forced_sparsity}
    \\\hfill\\
    \INIT
    \STATE Initialize $w_k^{(t)}$ so that $w_k^{(t)\top} \bar{\Var}[X_k^{(t)}] w_k^{(t)} = 1$ for all $t \in [T]$ and $k = 1, 2$. e.g.,
        \begin{equation}
            w_k^{(t)} \leftarrow  \mathbf{1} / \sqrt{\mathbf{1}^\top \bar{\Var}[X_k^{(t)}] \mathbf{1}}.
        \end{equation}
        and let
        \begin{equation}
            \bar\Sigma \leftarrow \bar{\Var}[w_1^{(1)\top} X_1^{(1)}, \dots, w_2^{(T)\top} X_2^{(T)}].
        \end{equation}
    \STATE  Initialize $\Sigma$ and $\Omega$ by
        \begin{equation}
            \Sigma \leftarrow \bar\Sigma + \lambda_\text{diag} I_{2T} \textand \Omega \leftarrow \Sigma^{-1}.
        \end{equation}
    \\\hfill\\
    \ITER
    \FOR{$\text{iter}$ in 1:$\text{iter}_\text{max}$}
        \STATE $\Sigma_{\text{last}} \leftarrow \Sigma$, $\Omega_{\text{last}} \leftarrow \Omega$
        \STATE $\Sigma, \Omega \leftarrow \texttt{P-gLASSO}(\Omega_\text{init}, \Sigma_\text{init}, \bar\Sigma, \Lambda, \text{iter}_\text{max}, \text{ths})$.
        \\\hfill\\
        \FOR{$k$ in 1:2 and $t$ in 1:$T$}
            \STATE $A \leftarrow \bar\Cov[X_k^{(t)}, (Y_l^{(s)}: (l,s) \neq (k,t))]$.
            \STATE $b \leftarrow (\Omega_{kl}^{(t,s)}: (l,s) \neq (k,t))$
            \IF{$Ab \neq \mathbf{0}$}
                \STATE $w_k^{(t)} \leftarrow \bar{\Var}(X_k^{(t)})^{-1} A b$
                \STATE $w_k^{(t)} \leftarrow w_k^{(t)} / \sqrt{w_k^{(t)\top} \bar{\Var}[X_k^{(t)}] w_k^{(t)}}$
            \ENDIF
        \ENDFOR
        \\\hfill\\
        \STATE $\bar \Sigma \leftarrow \bar{\Var}\left[w_1^{(1)\top} X_1^{(1)}, \dots, w_2^{(T)\top} X_2^{(T)}\right].$
        \IF{$\max(|\Sigma - \Sigma_{\text{last}}|) < \text{ths}$}
            \STATE \textbf{break}
        \ENDIF
    \ENDFOR
\end{algorithmic}
\label{alg:LaDynS}
\end{algorithm}

To update $\Omega$, we use
the P-gLASSO algorithm of \citet{mazumder2012graphical}, which is more efficient than the original gLASSO algorithm of \citet{friedman2008sparse}. The efficiency is attributed to P-gLASSO's flexibility with initial values, whereas gLASSO operates with a strict choice of initial $\hat{\Omega}$ ($\bar{\Sigma}^{-1}$ in case of \cref{eq:GLasso}). In \cref{alg:LaDynS}, the estimate $\hat{\Omega}$ from the past iteration serves as a warm start for the next iteration, so that we do not have to redo the entire paths from $\bar{\Sigma}^{-1}$ to $\hat{\Omega}$.
We can further reduce the computation cost by harnessing the banded sparse structure of $\Omega$ in \cref{fig:forced_sparsity}. 
In \cref{alg:pglasso}, we provide a modified P-gLASSO algorithm designed for the banded sparsity. The modified algorithm reduces the size of the LASSO sub-problem from $2T$ to $2d_\text{cross} + 2d_\text{auto}$ and the computational cost of a P-gLASSO iteration from $O(T^4 + T^3N)$ to $O(T((d_\text{cross}+d_\text{auto})^3+(d_\text{cross}+d_\text{auto})^2N)$.

\begin{algorithm}[h]
\caption{Modified \texttt{P-gLASSO} algorithm}
\begin{algorithmic}[1]
    \INPUT \hfill\\
    $\Omega_{\text{init}}, \Sigma_{\text{init}} \in \reals^{P \times P}$ : initial values, $\Sigma_{\text{init}} = (\Omega_{\text{init}})^{-1}$ \\
    $\bar{\Sigma} \in \reals^{P \times P}$: sample covariance matrix of a $P$-variate random variable\\
    $\Lambda \in [0,\infty]^{P \times P}$: sparsity penalty matrix \\
    $\text{iter}_\text{max} \in \nats_{+}$: maximum iteration \\
    $\text{ths} \in \reals_{+}$: threshold for convergence
    \OUTPUT $\Omega$ and $\Sigma = \Omega^{-1}$ which solves \cref{eq:GLasso}
    \INIT
    \STATE $\Sigma \leftarrow \Sigma_{\text{init}}, \Omega \leftarrow \Omega_{\text{init}}$
    \ITER
    \FOR{$\text{iter}$ in 1:$\text{max}_\text{iter}$}
      \STATE $\Sigma_{\text{last}} \leftarrow \Sigma$, $\Omega_{\text{last}} \leftarrow \Omega$
      \\\hfill\\
      \FOR{$p$ in 1:$P$}
        \STATE $D_k:$ the collection of $q$'s in $[P]$ s.t. $q \neq p$ and $\Lambda_{p,q} < \infty$ 
        \STATE $I_p:$ the collection of $q$'s in 1:$P$ s.t. $q \neq p$ and $\Lambda_{p,q} = \infty$ 
        \STATE (We notate the submatrix of a matrix $A \in \reals^{P \times P}$ of rows in $I \subset [d]$ and columns in $J \subset [P]$ by $A_{IJ}$. We moreover use $-p$ as a notation for $[d] \backslash \{p\}$ when it is used as a subscript of $A$.)
        \\\hfill\\
        \STATE $W = (\Omega_{-p,-p})^{-1}$ can be easily calculated by $\Sigma_{-p,-p} - \Sigma_{-p,p}\Sigma_{-p,p}/\Sigma_{p,p}$
        \\\hfill\\
        \STATE $\Sigma_{p,p} \leftarrow \bar{\Sigma}_{p,p} + \Lambda_{p,p}$
        \STATE $\Omega_{p,D_p}, \Omega_{D_p,p} \leftarrow \texttt{LASSO}(\Sigma_{p,p} \cdot W_{D_p,D_p}, -\bar{\Sigma}_{p,D_p}, \Lambda_{p,D_p})$ with an initial value $\Omega_{p,D_p}$.
        \STATE $\Omega_{p,I_p}, \Omega_{I_p,p} \leftarrow \mathbf{0}$
        \STATE $\Sigma_{p,-p}, \Sigma_{-p,p} \leftarrow - W_{:,D_p} \Omega_{D_p,p} \Sigma_{p,p}$
        \STATE $\Omega_{p,p} \leftarrow (1 - \Omega_{p,D_p}\Sigma_{D_p,p})/\Sigma_{p,p}$
        \STATE $\Sigma_{-p,-p} = W + \Sigma_{-p,p}\Sigma_{-p,p}/\Sigma_{p,p}$
      \ENDFOR
      \\\hfill\\
      \IF{$\max(|\Sigma - \Sigma_{\text{last}}|) < \text{ths}$}
        \STATE \textbf{break}
      \ENDIF
    \ENDFOR
\end{algorithmic}
\label{alg:pglasso}
\end{algorithm}

\subsection{Sensitivity of the Algorithm to Initial Values}
\label{app:sensitivity_analysis}

\begin{figure}[t!]

  \includegraphics[scale=0.6]{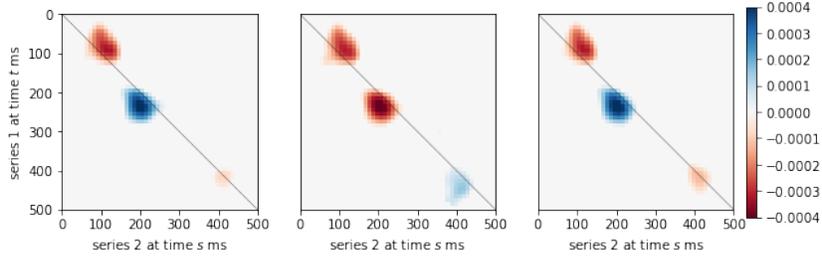}
  
  \caption{\sl {\bf LaDynS estimates $\hat{\Omega}_{12}$ by \cref{alg:LaDynS} for a given simulated dataset and three sets of initial values for $\beta_k^{(t)}$'s.} The dataset was generated as described in \cref{sec:simulation_from_SSM}, with \cref{fig:result_SSM}(a) showing the true ${\Omega}_{12}$. The three estimates $\hat{\Omega}_{12}$ are similar and close to ${\Omega}_{12}$, suggesting that \cref{alg:LaDynS} is not overly sensitive to initial values.
  }
  \label{fig:Omega_sensitivity}
\end{figure}

\begin{figure}
    \centering

    \raisebox{4.2cm}{\bf (a)}
    \includegraphics[scale = 0.6]{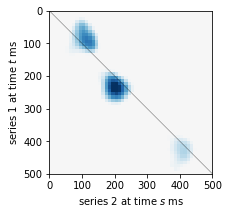}
    \quad
    \raisebox{4.2cm}{\bf (b)}
    \includegraphics[scale = 0.6]{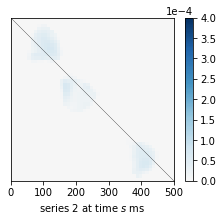}


    \caption{\sl {\bf (a) Mean and (b) entry-wise standard deviation of 1000 repeat $\Omega_{12}$ obtained by \cref{alg:LaDynS} for a fixed dataset and $1000$ different sets of initial $\beta_k^{(t)}$'s.}
    We used the same dataset as in \cref{fig:Omega_sensitivity}. 
    The mean estimated lead-lag effects in (a) are close to the true effects in \cref{fig:result_SSM}(a),  and the inferential uncertainty in (b) is small compared to the size of the effects in (a).
    }
    \label{fig:Omega_sensitivity_1000}
\end{figure}

We tested the sensitivity of \cref{alg:LaDynS} to the factor loadings $\beta_k^{(t)}$ initial values, under the simulation setting in \cref{sec:simulation_from_SSM}. Given a simulated dataset, we ran \cref{alg:LaDynS} starting from $1000$ different sets of initial values for $\beta_k^{(t)} \in \reals^{d_k}$ were randomly sampled independently across $k=1,2$ and $t \in [T]$ from a standard $d_k$-dimensional multivariate Gaussian distribution. \cref{fig:Omega_sensitivity} shows the LaDynS estimates $\hat{\Omega}_{12}$ from \cref{alg:LaDynS} for the same dataset and three sets of initial values. Although the three $\hat{\Omega}_{12}$'s have different values, they yield similar inference for lead-lag relationships between the two time-series. The most obvious differences are the signs of the estimated precision entries, but these have no implication on lead-lag inferences because 
the signs of the factor loadings $\beta_k^{(t)}$ and the latent precision matrix $\Omega$ are not identifiable. \cref{fig:Omega_sensitivity_1000}(a) confirms that the lead-lag inferences are consistent with the simulation setting in \cref{sec:simulation_from_SSM} and that the inferential uncertainty in \cref{fig:Omega_sensitivity_1000}(b) is small compared to the size of the effect in \cref{fig:Omega_sensitivity_1000}(a).

\begin{figure}
    \centering

    \raisebox{3.7cm}{\bf (a)}
    \includegraphics[scale=0.6]{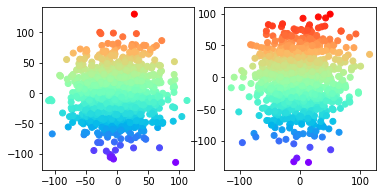}

    \raisebox{3.7cm}{\bf (b)}
    \includegraphics[scale=0.6]{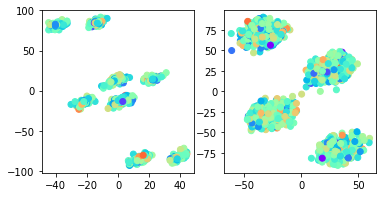}
    
    

    \caption{\sl {\bf Scatterplots of 2D random projections of (a) $1000$ different initial values of $\beta_k^{(t)}$'s and (b) corresponding fitted values after applying \cref{alg:LaDynS} to the dataset used in \cref{fig:Omega_sensitivity}.} The same projections were used for final and initial values. The left and right panels correspond to the projected $\beta_1$ and $\beta_2$, respectively. 
    The distributions of initial values are random, by design. Fitted values converge to the same locations, further confirming that \cref{alg:LaDynS} 
    is not sensitive to initial values.
    }
    \label{fig:beta_sensitivity_1000}
\end{figure}

Next, we checked that the estimated factor loadings were similar across the $1000$ runs of \cref{alg:LaDynS} with different initial values. Because they are $(d_1 + d_2) \times T$-dimensional, we projected them on an arbitrary $2$-dimensional subspace to visualize them. 
\cref{fig:beta_sensitivity_1000} shows the projected initial and final values of the factor loadings
across the $1000$ runs of \cref{alg:LaDynS}.
The initial values in \cref{fig:beta_sensitivity_1000}(a) 
are randomly distributed, by design, and converge to a distribution that is consistent across repeat simulations.
\cref{fig:beta_sensitivity_1000} is consistent with our observation in \cref{fig:Omega_sensitivity} that the main source of the sensitivity to initial values  are the signs of the precision entries in the three epochs of lead-lag relationships. Indeed, because the signs for each epoch are not identifiable, we expect $2^3 = 8$ local minima in \cref{eq:penalized_likelihood}, where each minimum corresponds to a set of possible signs for the three epochs. Therefore, we expect see at most $8$ clusters in \cref{fig:beta_sensitivity_1000}(b) (at most because clusters can overlap one another). The fitted factor loadings are not exactly equal within a cluster across repeat simulations because we optimize the likelihood \cref{eq:penalized_likelihood} numerically.

%% file: appendices/c_supp_simulation.tex
\section{Supplementary Results on Simulated Data} \label{app:supp_simulation}

\subsection{Simulated datasets with amplitude correlation but no coherence} \label{app:simulation_no_coh}
Here we illustrate that LaDynS can discover lead-lag relationships in amplitude where frequency-domain analyses based on coherence or phase locking cannot.

\begin{figure}[h!]
    \centering

    \raisebox{4.2cm}{\bf (a)}
    \includegraphics[scale=0.59]{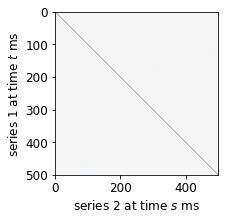}
    \quad
    \raisebox{4.2cm}{\bf (b)}
    \includegraphics[scale=0.6]{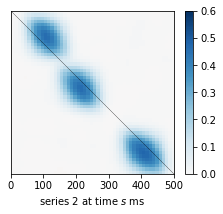}
    \quad
    \raisebox{4.2cm}{\bf (c)}
    \includegraphics[scale=0.6]{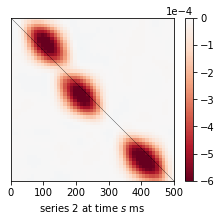}
  
  
  \caption{\sl {\bf
  {\bf (a)} Cross-coherence matrix $C_{12}$, {\bf (b)} cross-covariance matrix $\Sigma_{12}$ and {\bf (c)} cross-precision matrix $\Omega_{12}$ of two latent time-series simulated as in \cref{app:simulation_no_coh}}. Lead-lag relationships between amplitudes of the simulated time series are present, but there is no coherence between their phases.
  }
  \label{fig:Coh_Sigma_Omega_woc}
\end{figure}

We modified the simulation setting of \cref{sec:simulation_from_SSM} to generate time series that have lead-lag relationships in amplitude but no coherence.  
We generated two sets of multi-variate LFP time series using \cref{eq:nonstationary_lagged_mixture_model}, except that $L_1$ and $L_2$ were randomly phase-shifted from $L_{0,j}$: 
\begin{equation} \label{eq:woc_model}
    L_k^{(t)} = \sum_{j=1}^3 \beta_{kj} \cdot e^{i \omega_{kj}} \cdot L_{0,j}^{(t-\tau_{kj})} + \eta_k^{(t)}, ~\text{for}~ k=1,2,
\end{equation}
where $i$ in the exponential function is the complex unit, and $\omega_{kj}$ is a random phase shift sampled from Uniform distribution $U[0,2\pi]$. One simulation dataset consisted of $1000$ trials, and the coherence and amplitude correlation at $f_0=18$ Hz was studied. 
We filtered the simulated LFPs at frequency $f_0$ as in \cref{sec:simulation_from_SSM}. Let $(\tilde{X}_{1}^{(t)}, \tilde{X}_{2}^{(t)})$ be the filtered data downsampled to $100$ Hz; these data are complex-variate, of which the arguments and absolute values are the oscillatory phases and amplitudes of $L_k^{(t)}$ at frequency $f_0$, for $k=1,2$ respectively. That is, the beta power envelope $X_{k}^{(t)}$ is the absolute value of $\tilde{X}_k^{(t)}$. In \cref{sec:simulation_result}, we recovered the true latent factor $(Z_1^{(t)}, Z_2^{(t)})$ as
\begin{equation*}
    Z_k^{(t)} = w_k^{(t)\top} X_k^{(t)} ~~\text{for}~~ t=1,\dots,T, ~~k=1,2,
\end{equation*}
based on \cref{eq:canonical_variable} and known factor loadings $(\beta_1^{(t)}, \beta_2^{(t)})$. 
Here we also define 
\begin{equation*}
    \tilde Z_k^{(t)} = w_k^{(t)\top} \tilde X_k^{(t)} ~~\text{for}~~ t=1,\dots,T, ~~k=1,2.
\end{equation*}
The cross-coherence $C_{12}^{(t,s)}$ between $\tilde{Z}_1^{(t)}$ and $\tilde{Z}_2^{(s)}$ is the population coherence between $\tilde{X}_1^{(t)}$ and $\tilde{X}_2^{(s)}$, whereas the cross-correlation $\Sigma_{12}^{(t,s)}$ between $Z_1^{(t)}$ and $Z_2^{(s)}$ is the population amplitude correlation.
\cref{fig:Coh_Sigma_Omega_woc} shows the estimates of $C_{12}$, $\Sigma_{12}$ and the amplitude cross-precision $\Omega_{12}$ based on the average of $200$ repeats, as in \cref{sec:simulation_result}.
We can verify that there is no cross-coherence between the two time series and that the amplitude cross-correlation is as in \cref{sec:simulation_from_SSM}. 

\begin{figure}[h!]
  \centering
  \includegraphics[scale=0.6]{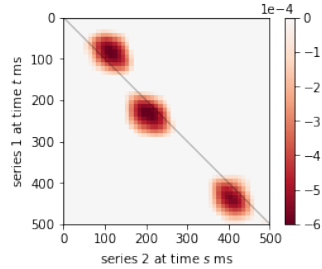}

  \caption{\sl {\bf LaDynS cross-precision matrix estimate $\hat{\Omega}_{12}$ to the simulated data without cross-coherence.}
  LaDynS can discover lead-lag relationships in amplitude where frequency-domain analyses based on coherence or phase locking cannot. 
  }
  \label{fig:Omega_est_woc}
\end{figure}

To verify that LaDynS can estimate the lead-lag relationships in amplitude,  we applied LaDynS to a simulated beta power envelope dataset $(X_1^{(t)}, X_2^{(t)})$.
\cref{fig:Omega_est_woc} shows the LaDynS estimate of $\Omega_{12}$. 
Although there is no coherence between the two time series, the LaDynS estimate recovers the true cross-precision in \cref{fig:Coh_Sigma_Omega_woc}(c), when coherence-based frequency-domain methods cannot find a significant coherence at any pair of time points. 

\subsection{Analysis of simulated datasets with known canonical correlation matrix} \label{app:simulation_from_model}

\begin{figure}[t!]
   \centering 
   \raisebox{4.2cm}{\bf (a)}
    \includegraphics[scale=0.6]{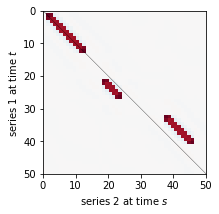}
    \quad
    \raisebox{4.2cm}{\bf (b)}
    \includegraphics[scale=0.6]{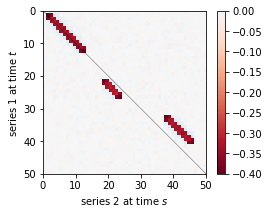}
    

    \raisebox{4.2cm}{\bf (c)}
    \includegraphics[scale=0.6]{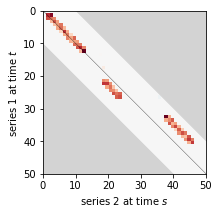}
    \quad
    \raisebox{4.2cm}{\bf (d)}
    \includegraphics[scale=0.6]{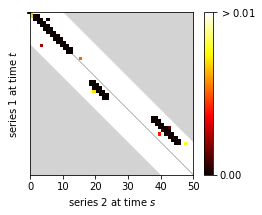}
    \quad
    \raisebox{4.2cm}{\bf (e)}
    \includegraphics[scale=0.6]{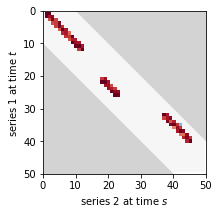}
    
    
    \caption{\sl 
    {\bf Output and inference of LaDynS applied to one simulated dataset from the LaDynS model.}
    {\bf (a)} True cross-precision matrix $\Omega_{12}$, for the connectivity scenario described in \cref{app:simulation_from_model} with $r=0.4$.
    {\bf (b)} Average over $60$ simulation datasets of LaDynS de-sparsified precision estimates $\tilde\Omega_{12}$. There is a good match to the true $\Omega_{12}$ in (a).
    {\bf (c)} Cross-precision estimate $\hat\Omega_{12}$ for one simulated dataset. It matches (a) up to random error.
    {\bf (d)} Permutation bootstrap p-values for the de-sparsified estimate $\tilde \Omega_{12}$.
    {\bf (e)} Discovered non-zero cross-precision estimates by the BH procedure at nominal FDR $5$\%. The cluster-wise p-values of the three discovered clusters by the excursion test were all smaller than $0.5\%$. All the panels but (d) share the same color bar in (b).}
    \label{fig:result_model}
    
\end{figure}

To further examine the properties of LaDynS and associated inference procedures, we present the results of another simulation study. 
One simulated dataset consisted of $N = 1000$ i.i.d. vector time-series $X_1$ and $X_2$ of dimensions $d_1 = d_2 = 25$ and durations $T=50$, simulated from \cref{eq:pCCA_obs_over_time,eq:pCCA_latent_over_time}.
The latent time series $Z_1$ and $Z_2$ in \cref{eq:pCCA_obs_over_time,eq:pCCA_latent_over_time} had zero mean vectors and covariance matrix $\Sigma = \Omega^{-1}$, with
\begin{equation}\label{eq:true_omega_sim}
    \Omega = \begin{bmatrix}
   ( \Sigma_{0,1} +\lambda I_T)^{-1} & \Omega_{12} \\
    \Omega_{12}^\top&( \Sigma_{0,2} +\lambda I_T)^{-1} \\
    \end{bmatrix},
\end{equation}
where $\Omega_{12}$ was the cross-precision matrix of interest.
The elements of the auto-precision matrices were simulated from the squared exponential function: 
\begin{equation}\label{eq:sq_exp_cov}
    \Sigma_{0,k}^{(t,s)} = \exp\left(-c_{0,k}(t-s)^2\right), \,\, k=1,2
\end{equation}
with $c_{0,1} = 0.148$ and $c_{0,2} = 0.163$ chosen to match the LFPs autocorrelations in the experimental dataset. The diagonal regularizer $\lambda I_T$ was added to ensure that $\Sigma_{0,1}$ and $\Sigma_{0,2}$ were invertible, and 
we set $\lambda=1$. 
%
For $\Omega_{12}$, we considered the connectivity scenario depicted in \cref{fig:result_model}(a), where the two latent times series connected in three epochs, the first with no latency, the second with series 2 preceding series 1, and the third with series 1 preceding series 2.
We accordingly set the cross-precision matrix elements to
\begin{equation}\label{eq:simulated_Omega_cross}
    \Omega_{12}^{(t, s)} = \begin{cases}
    -r, & \text{if } (t,s) \text{ is colored red}, \\
    0,& \text{elsewhere},
    \end{cases}
\end{equation}
where $r$ measured the intensity of the connection.
Finally, we rescaled $\Sigma$ to have diagonal elements equal to one. 

Once the latent time series $Z_1$ and $Z_2$ were generated, we simulated a pair of observed time series according to
\begin{equation}\label{eq:simulated_obs}
  X_k^{(t)} = Y_k^{(t)} - \beta_k^{(t)} w_k^{(t)\top} \left(Y_k^{(t)} - \Exp[Y_k^{(t)}]\right) + \beta_k^{(t)} Z_k^{(t)},
\end{equation}
for $k=1,2$ and $t=1,\dots,T$, where $Y_1^{(t)}$ and $Y_2^{(t)}$ were uncorrelated baseline time series, 
$\beta_k^{(t)}$ were factor loadings that change smoothly over time, $w_k^{(t)}$ were canonical weights that satisfy 
the relationship with $\beta_k^{(t)}$ in \cref{eq:genvar_solution},
and $\Exp[Y_k^{(t)}]$ was the mean of $Y_k^{(t)}$, for $k=1,2$. We subtracted $\beta_k^{(t)} w_k^{(t)\top} \left(Y_k^{(t)} - \Exp[Y_k^{(t)}]\right)$ to ensure that $X_k^{(t)}$ had canonical correlation matrix $\Sigma$ and the same mean as $Y_k^{(t)}$.
Note that, unlike in \cref{sec:simulation_from_SSM}, the canonical correlation matrix was known exactly. 
We took $Y_1^{(t)}$ and $Y_2^{(t)}$ to be the two multivariate time-series of neural recordings analyzed in \cref{sec:experimental_data}, which we permuted to remove all cross-correlations. 
To reduce temporal auto-correlations, we added space-correlated white noise to the baseline time series. The amount of noise was set to be comparable to the diagonal regularization $\lambda I_T$ introduced in \cref{eq:true_omega_sim}.
Finally we set $\beta_k \in \mathbb{R}^{d_k \times T}$ to be the factor loadings estimated in \cref{sec:experimental_data}. 
The resulting latent time series $\beta_k^{(t)} Z_k^{(t)}$ and noise baseline vector $Y_k^{(t)}$ in \cref{eq:simulated_obs} had comparable scales and auto-correlations by construction, for $k=1,2$, 
to the experimental data in \cref{sec:experimental_data}. 

\smallskip
\noindent {\it LaDynS estimation details. \,\,}\label{app:simulation_estimation}
For this simulation, we did not need to regularize the diagonal of $\Omega$, because the simulated time series were not smooth, and the resulting $\hat{\Sigma}$ was invertible without the regularization. 
Hence we set $\lambda_\text{diag} = 0$.
The other hyperparameters were set to  
$d_\text{auto} = d_\text{cross} = 10$, and $\lambda_\text{auto} = 0$. The penalty on the cross-correlation elements, $\lambda_\text{cross}$, was automatically tuned at every repeat of simulation to control false discoveries (see \cref{sec:LaDynS}).

\smallskip
\noindent {\it Results.\,\,}
\Cref{fig:result_model}(c) displays the LaDynS cross precision estimate $\hat\Omega_{12}$ fitted to one dataset simulated under the connectivity scenario depicted in \cref{fig:result_model}(a),
with connection strength $r=0.4$ in \cref{eq:simulated_Omega_cross}. 
\Cref{fig:result_model}(d) shows the permutation bootstrap p-values for the entries of the desparsified cross-precision estimate $\tilde\Omega_{12}$ (\cref{eq:p_prec} with permutation bootstrap simulation size $B = 200$; see \cref{sec:inference}). Small p-values concentrate near the locations of true non-zero cross-precision entries and are otherwise scattered randomly. 
We applied first the BH procedure with target FDR $5\%$ (\cref{sec:inference}) 
and subsequently the excursion test at the 5\% significance level to all discovered clusters. The significant clusters 
($p < 0.005$) are plotted in \cref{fig:result_model}(e). 
They match approximately the true clusters in \cref{fig:result_model}(a), although they
exhibit random variability. To average this random variability out, we
estimated $\Omega_{12}$ for each of 60 simulated datasets, and plotted their average in \cref{fig:result_model}(b). The
average LaDynS estimate is a close match to the true cross-precision matrix in \cref{fig:result_model}(a).

\begin{figure}[t!]
  \centering
  
    \includegraphics[scale=0.6]{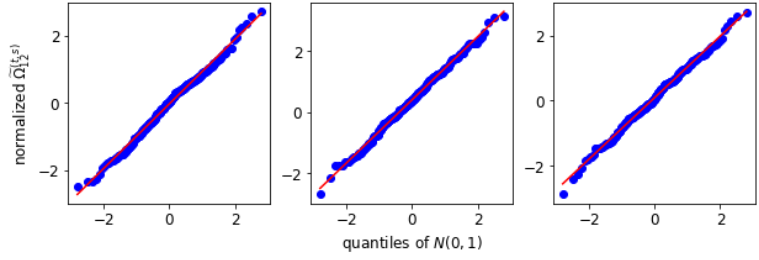}

  \caption{
  \sl {\bf Null distributions of three representative entries of $\tilde{\Omega}_{12}^{(t,s)} / \sqrt{\hat{\Var}[\tilde{\Omega}_{12}^{(t,s)}]}$} obtained from $R = 60$ simulated datasets (\cref{app:simulation_from_model}), compared to the standard Gaussian distribution via QQ-plots. There is good agreement. 
  }
  \label{fig:qq_dspr}
\end{figure}

\begin{figure}[t!]
    \centering 
    \raisebox{4cm}{\bf (a)}
    \includegraphics[scale=0.6]{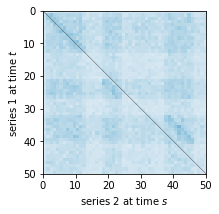}
    \quad
    \raisebox{4cm}{\bf (b)}
    \includegraphics[scale=0.6]{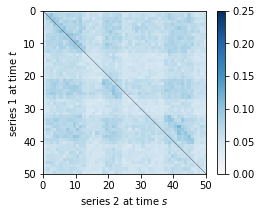}
    \quad
    \raisebox{4cm}{\bf (c)}
    \includegraphics[scale=0.6]{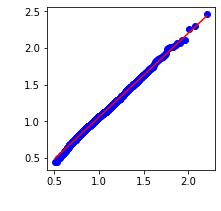}
  
  
    \caption{{\bf Standard deviations of desparsified precision elements.} \sl Variance obtained {\bf(a)} from samples from the ground-truth generative multiset pCCA model and {\bf (b)} from permutation bootstrapped samples. {\bf (c)} F-statistics of ratios between the two variances for null entries of $\Omega_{12}$, showing good agreement.}
  \label{fig:sd_sim_vs_perm}
\end{figure}

\smallskip
\noindent{\it Normal approximation for the p-values in \cref{eq:p_prec}. \,\, }
We investigated the validity of the Normal assumption by comparing the empirical distribution of $R=60$ repeat estimates $\tilde{\Omega}_{12}^{(t,s)} / \sqrt{\hat{\Var}[\tilde{\Omega}_{12}^{(t,s)}]}$ (\cref{eq:CLT_on_dspr_LaDynS}) to the standard normal distribution using QQ-plots. 
\cref{fig:qq_dspr} shows QQ-plots for three randomly chosen representative time pairs $(t,s)$ that are such that $\Omega_{12}^{(t,s)} = 0$, which validates the normal assumption. 
%
We further checked the validity of the permutation bootstrap variance estimates $\hat\Var[\tilde\Omega_{12}^{(t,s)}]$,
shown in \cref{fig:sd_sim_vs_perm}(b),
by comparing them to the empirical variances of $R=60$ estimates $\tilde{\Omega}_{12}^{(t,s)}$, shown in \cref{fig:sd_sim_vs_perm}(a).
There is good agreement for the entries that have precision value zero, $\Omega_{12}^{(t,s)} = 0$. 
\cref{fig:sd_sim_vs_perm}(c) further displays the Q-Q plot of the repeat ratios of permutation bootstrap over empirical estimates of $\Var[\Omega_{12}^{(t,s)}]$ for these entries, with $F(B-1, R-1)$ being the reference distribution. The good agreement suggests that the bootstrap estimate of $\Var[\Omega_{12}^{(t,s)}]$ is reliable.

\begin{figure}[t!]
    \centering 
    \raisebox{4cm}{\bf (a)}
    \includegraphics[scale=0.6]{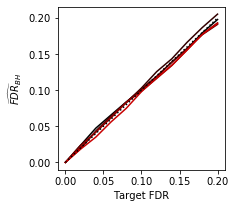}
    \quad
    \raisebox{4cm}{\bf (b)}
    \includegraphics[scale=0.6]{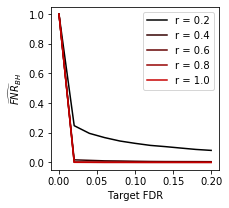}
  
  
  \caption{\sl {\bf False Discovery Rate control for LaDynS' inference.}
  {\bf (a)} Estimated false discovery rate and
  {\bf (b)} false non-discovery rate 
  for target $\text{FDR} \in [0, 20]\%$, under the connectivity scenario in \cref{fig:result_model}(a), for connectivity intensities $r = 0.2$, $0.4$, $0.6$, $0.8$ and $1.0$ in \cref{eq:simulated_Omega_cross}. 
  The dotted line is a (0,1) line. 
  }
  \label{fig:fdr_fnr_model}
\end{figure}

\smallskip
\noindent{\it FDR control. \,\, }
\cref{fig:fdr_fnr_model} shows that estimated and target FDR values match for a range of connection strengths ($r$ in \cref{eq:simulated_Omega_cross}).
In addition, the FNR is very low for target FDRs larger than 2\%.

\begin{figure}[t!]
    \centering
    \raisebox{4cm}{\bf (a)}
    \includegraphics[scale=0.6]{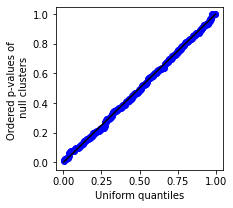}
    \quad
    \raisebox{4cm}{\bf (b)}
    \includegraphics[scale=0.6]{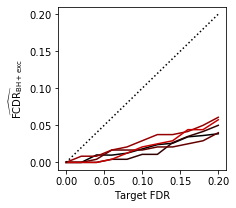}
    \quad
    \raisebox{4cm}{\bf (c)}
    \includegraphics[scale=0.6]{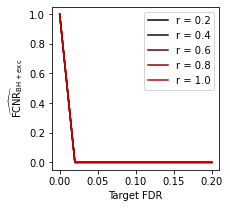}

  
  \caption{\sl {\bf Performance of cluster-wise inference after excursion test.}
  {\bf (a)} Q-Q plot of excursion test p-values from null clusters versus $\mathrm{Uniform}[0,1]$ distribution, as expected of a valid test.
  {\bf (b)} false cluster discovery rates and
  {\bf (c)} false cluster non-discovery rate of BH at target FDR $10 \%$ followed by excursion test at significance level $\alpha \in [0, 0.10]$ to identify non-zero partial correlations, under the connectivity scenario in \cref{fig:result_SSM}(a), for the simulated range of connectivity intensities.
  }
  \label{fig:qq_fcdr_fcnr_exc}
\end{figure}

\smallskip
\noindent{\it Excursion test. \,\, } 
The next step was to apply the excursion test to each cluster discovered by the BH procedure at target FDR $5\%$. As a check on the validity of this test, \cref{fig:qq_fcdr_fcnr_exc}(a) shows the quantiles of null p-values versus the theoretical quantiles of the uniform distribution on $[0,1]$. They match perfectly so the excursion test is reliable. 
Finally, \cref{fig:qq_fcdr_fcnr_exc}(b,c) display the cluster-wise FDR and FNR (FCDR and FCNR) for the BH procedure followed by the excursion test. 
The estimated FCDRs (\cref{fig:qq_fcdr_fcnr_exc}(b)) are small throughout the tested range of nominal FDR values and connectivity intensities. The estimated FCNRs (\cref{fig:qq_fcdr_fcnr_exc}(c)) are zero for nominal FDRs greater than 2\%, which means that all connectivity epochs in \cref{fig:result_model}(a) were discovered by our methods in the simulated dataset.

  
  

%% file: appendices/d_supp_experiment.tex
\section{Supplementary Figures for the Experimental Data Analysis in Section~\ref{sec:experimental_data}} 

\label{app:supp_experiment}

\begin{figure}[h!]
    \centering
    \raisebox{5cm}{\bf (a)}
    \includegraphics[scale=0.5]{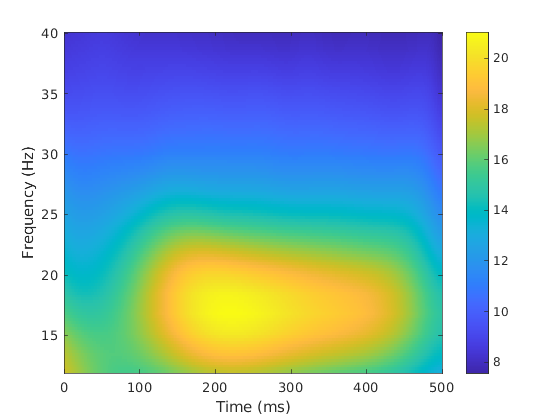}
    \quad
    \raisebox{5cm}{\bf (b)}
    \includegraphics[scale=0.5]{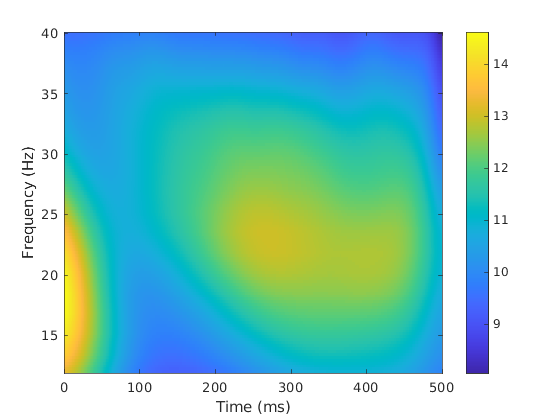}
    
    
    \caption{\sl {\bf Averaged spectrograms across trials and electrodes in (a) V4 and (b) PFC as functions of experimental time}, cropped at frequency between $12$ Hz and $40$ Hz to focus on the beta band oscillations.}
    \label{fig:spectrogram_Smith}
\end{figure}

\begin{figure}[h!]
  \centering
  \includegraphics[scale=0.6]{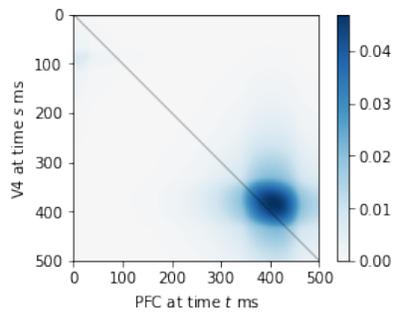}

  \caption{\sl {\bf Latent cross-correlation matrix estimate $\hat{\Sigma}_{12}$.} }
  \label{fig:Sigma_Smith}
\end{figure}

\begin{figure}[h!]
    \centering

    \raisebox{4.2cm}{\bf (a)}
    \includegraphics[scale=0.6]{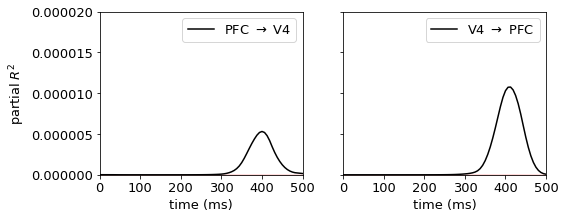}
    \quad

    \raisebox{4.2cm}{\bf (b)}
    \includegraphics[scale=0.6]{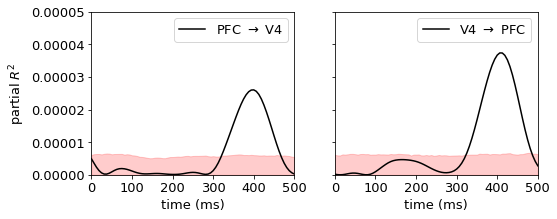}
    \quad

  
  
  \caption{\sl {\bf Estimated partial $R^2$ from locally stationary state-space model} based on (a) the LaDyns precision estimate $\hat{\Omega}$ and (b) desparsified estimate $\tilde{\Omega}$ for V4 $\rightarrow$ PFC and PFC $\rightarrow$ V4. The pink shaded areas are the 95th percentiles of null partial $R^2$ under independence between V4 and PFC.}
  \label{fig:pR2_results_What_dspr}
\end{figure}

\begin{figure}[h!]

  \includegraphics[scale=0.6]{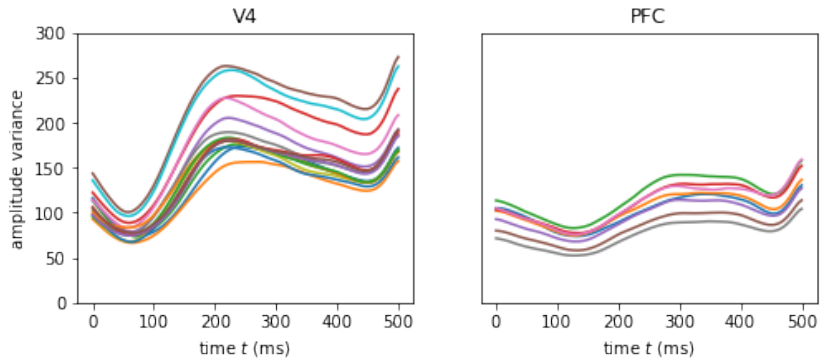}
  \caption{\sl {\bf Estimated variances of beta amplitudes at active electrodes in V4 and PFC as functions of time}. 
  The active electrodes were those with factor loading values larger than 75\% of the maximal value at experimental time $400$ ms (\cref{fig:beta_results} shows the factor loadings at $400$ ms over the electrode arrays). There were $19$ active electrodes in V4 and $16$ in PFC. 
  }
  \label{fig:fnorm_active_Smith}
\end{figure}